\documentclass[12pt]{article}             
\usepackage{feynmp}                       

\usepackage[dvips]{color}                 
\usepackage{graphicx}                     

\newcommand{\comment}[1]{}
\newcommand{\note}[1]{}

\newcommand{\ia}{{\"{\i}}}   
\newcommand{\absatz}{\vspace{2ex}\noindent}

\newcommand{\ii}{\mathrm{i}}

\newcommand{\tr}{\mathrm{tr}}
\newcommand{\T}{\mathrm{T}}

\newcommand{\kv}{\vec{k}}
\newcommand{\pv}{\vec{\,\!p}\!\:{}}
\newcommand{\qv}{\vec{\,\!q}\!\:{}}

\newcommand{\mpi}{m_\pi}
\newcommand{\fpi}{f_\pi}
\newcommand{\MeV}{\mathrm{MeV}}
\newcommand{\fm}{\mathrm{fm}}

\newcommand{\de}{\partial}
\newcommand{\dev}{\vec{\de}}

\newcommand{\calC}{\mathcal{C}}\newcommand{\calD}{\mathcal{D}}
\newcommand{\calE}{\mathcal{E}}\newcommand{\calF}{\mathcal{F}}
\newcommand{\calI}{\mathcal{I}}\newcommand{\calK}{\mathcal{K}}

\newcommand{\calP}{\mathcal{P}}\newcommand{\calR}{\mathcal{R}}
\newcommand{\calT}{\mathcal{T}}\newcommand{\calY}{\mathcal{Y}}

\begin{document}

\begin{fmffile}{quafeyn}
\fmfset{curly_len}{2mm}
\fmfset{dash_len}{1.5mm}
\fmfset{wiggly_len}{3mm}
\newcommand{\feynbox}[2]{\mbox{\parbox{#1}{#2}}}
\newcommand{\fs}{\scriptstyle} 
\newcommand{\hq}{\hspace{0.5em}}
\newcommand{\hqm}{\hspace{-0.25em}}

\fmfcmd{vardef ellipseraw (expr p, ang) = save radx; numeric radx; radx=6/10
  length p; save rady; numeric rady; rady=3/10 length p; pair center;
  center:=point 1/2 length(p) of p; save t; transform t; t:=identity xscaled
  (2*radx*h) yscaled (2*rady*h) rotated (ang + angle direction length(p)/2 of
  p) shifted center; fullcircle transformed t enddef;
  style_def ellipse expr p=
                shadedraw ellipseraw (p,0);
                enddef; }

%

\begin{titlepage}
\begin{flushright}
  nucl-th/9907077\\ DOE/ER/40561-33-INT98\\NT@UW-99-3 \\ 15th July 1999 \\
\end{flushright}
\vspace*{1.5cm}
\begin{center}
  
  \LARGE{\textbf{Quartet $\mathrm{S}$ Wave Neutron Deuteron Scattering in
      Effective Field Theory}}

\end{center}
\vspace*{1.0cm}
\begin{center}
  
  \textbf{Paulo F.\ Bedaque${}^{a,\,}$\footnote{Email:
      bedaque@phys.washington.edu}} and \textbf{Harald W.\ 
    Grie\3hammer${}^{b,\,}$\footnote{Email: hgrie@phys.washington.edu}}
  
  \vspace*{0.2cm}
  
  \emph{${}^a$Institute for Nuclear Theory,
    University of Washington,\\
    Box 351 550, Seattle, WA 98195-1550, USA\\
    ${}^b$Nuclear Theory Group,
    Department of Physics, University of Washington,\\
    Box 351 560, Seattle, WA 98195-1560, USA} \vspace*{0.2cm}

\end{center}

\vspace*{2.0cm}

\begin{abstract}
  The real and imaginary part of the quartet $\mathrm{S}$ wave phase shift in
  $nd$ scattering (${}^4\mathrm{S}_{3/2}$) for centre-of-mass momenta of up to
  $300\;\MeV$ ($E_\mathrm{cm}\approx70\;\MeV$) is presented in effective field
  theory, using both perturbative pions and a theory in which pions are
  integrated out. As available, the calculation agrees with both experimental
  data and potential model calculations, but extends to a higher, so far
  untested momentum r\'egime above the deuteron breakup point. A Lagrangean
  more feasible for numerical computations is derived.
\end{abstract}
\vskip 1.0cm
\noindent
\begin{tabular}{rl}
Suggested PACS numbers:& 11.80.J, 21.30.-x, 21.45.+v, 25.10.+s, 25.40.Dn\\[1ex]
Suggested Keywords: &\begin{minipage}[t]{10cm}
                    effective field theory, nucleon-deuteron
                    scattering, three-body systems
                    \end{minipage}
\end{tabular}
\end{titlepage}

\setcounter{page}{2} \setcounter{footnote}{0} \newpage

%

\section{Introduction}
\label{sec:intro}

Effective field theory (EFT, for an introduction, see e.g.\ 
\cite{LepageQEDlecture}) methods are largely used in many branches of physics
where a separation of scales exists. In low energy nuclear systems, the two
well separated scales are, on one side, the low scales of the typical momentum
of the process considered and the pion mass, and on the other side the higher
scales associated with chiral symmetry and confinement. This separation of
scales was explored with great success in the mesonic sector (Chiral
Perturbation Theory~\cite{leut,Weinberg79}) and in the one baryon sector
(usually by using Heavy Baryon Chiral Perturbation
Theory~\cite{Gasseretal,JenkinsManohar}), producing a low energy expansion of a
variety of observables. It provided for the first time a systematic, rigorous
and model independent (meaning, independent of assumptions about the
non-perturbative QCD dynamics) description of strongly interacting particles. A
large amount of work was devoted recently to the application of EFT methods to
systems containing two nucleons. In the three nucleon sector, some progress was
also made in the very low energy range and at low orders where pions do not
have to be included explicitly~\cite{Stooges,Stooges2}. In this article, we investigate the quartet
$\mathrm{S}$ wave of neutron-deuteron scattering using the scheme advanced in
\cite{KSW} in which pions are included perturbatively.

The original suggestion of how to extend EFT methods to systems containing two
or more nucleons is due to Weinberg~\cite{Weinberg} who noticed that below the
$\Delta$ production scale, only nucleons and pions need to be retained as the
infrared relevant degrees of freedom of low energy QCD.  Because at these
scales the momenta of the nucleons are small compared to their rest mass, the
theory becomes non-relativistic at leading order in the velocity expansion,
with relativistic corrections systematically included at higher orders. The
most general chirally invariant Lagrangean consists hence of contact
interactions between non-relativistic nucleons, and between nucleons and pions.
The interactions involving pions are severely restricted by chiral invariance.
As such, the theory is an extension to the many nucleon system of Chiral
Perturbation Theory and its counterpart in the one nucleon sector, Heavy Baryon
Chiral Perturbation Theory. Like in its cousins, all short distance physics --
quarks and gluons, resonances like the $\Delta$ or $\sigma$ -- is integrated
out into the coefficients of the low energy Lagrangean. In principle, these
constants could be derived by solving QCD, for example, on the
lattice~\cite{Rebbi}.  They can also be determined through models of the short
distance physics, like resonance saturation. The most common way to determine
those constants, though, is by fitting them to experiment.

Because its Lagrangean consists of infinitely many terms only restricted by
symmetry, an effective field theory may at first sight suffer from lack of
predictive power. Indeed, as part of its formulation, predictive power is
ensured only by establishing a power counting scheme, i.e.\ a way to determine
at which order in a momentum expansion different contributions will appear, and
keeping only and all the terms up to a given order. The dimensionless, small
parameter on which the expansion is based is the typical momentum $Q$ of the
process in units of the scale $\Lambda$ at which the theory is expected to
break down, with estimates ranging from $\Lambda\approx 300$ to
$800\;\MeV$~\cite{INTWorkshopSummary} in the two body system. Assuming that all
contributions are of natural size, i.e.\ ordered by powers of $Q$, the
systematic power counting ensures that the sum of all terms left out when
calculating to a certain order in $Q$ is smaller than the last order retained,
allowing for an error estimate of the final result.

In systems involving two or more nucleons, establishing such a power counting
is complicated by the fact that unnaturally large scales have to be
accommodated, so that some coefficients in the Lagrangean may not be of natural
size and hence possibly jeopardise power counting: Given that the typical low
energy scale in the problem should be the mass of the pion as the lightest
particle emerging from QCD, fine tuning seems to be required to produce the
large scattering lengths in the ${}^1\mathrm{S}_0$ and ${}^3\mathrm{S}_1$
channels ($1/a^{{}^1\mathrm{S}_0}=-8.3\;\MeV,\;
1/a^{{}^3\mathrm{S}_1}=36\;\MeV$).  Since there is a bound state in the
${}^3\mathrm{S}_1$ channel with a binding energy $B=2.225\;\MeV$ and hence a
typical binding momentum $\gamma=\sqrt{M B}\simeq 46 \mathrm{MeV}$ well below
the scale $\Lambda$ at which the theory should break down, it is also clear
that at least some processes have to be treated non-perturbatively in order to
accommodate the deuteron.

A way to incorporate this fine tuning into the power counting was suggested by
Kaplan, Savage and Wise \cite{KSW}. In it, renormalisation using an unusual
subtraction scheme (called Power Divergence Subtraction, PDS) moves a somewhat
arbitrary amount of the short distance contributions from loops to counterterms
and makes precise cancellations manifest which arise from fine tuning.  Power
counting then becomes straightforward, but physical observables are of course
independent of the renormalisation scale or cut-off chosen. The leading order
contribution to nucleons scattering in an $\mathrm{S}$ wave comes from four
nucleon contact interactions and is summed geometrically to all orders to
produce the shallow real and virtual bound states. Pion interactions and
contact operators containing derivatives are suppressed by additional powers of
$\mpi$ and $Q$, respectively. In the pion contributions, the exchange of one
instantaneous pion is the leading piece, and pion ladders and radiative pions
are suppressed further. Dimensional regularisation is chosen to explicitly
preserve the systematic power counting as well as all symmetries (esp.~chiral
invariance) at each order in every step of the calculation. At leading (LO),
next-to-leading order (NLO) and often even NNLO in the two nucleon system, this
approach also allows in general for simple, closed answers whose analytic
structure is readily asserted ~\cite{N2LO}. This theory has been put to
extensive tests at NLO and NNLO in the two body system, giving for the first
time analytic answers to many deuteron properties such as electromagnetic form
factors~\cite{em}, scalar and tensor electromagnetic
polarisabilities~\cite{pola}, Compton scattering~\cite{Compton}, $np\to
d\gamma$ both parity violating~\cite{Pbreak} and conserving~\cite{Pcons}, the
deuteron anapole moment~\cite{ana}, charge dependence and charge symmetry
breaking~\cite{iso}. The expansion parameter is found to be of the order of
$\frac{1}{3}$, so that NLO calculations are accurate to about $10\%$, and NNLO
calculations to about $3\%$. In all cases, experimental agreement is within the
estimated theoretical uncertainties, and in some cases, previously unknown
counterterms could be determined.

Even if calculations of nuclear properties were possible starting from the
underlying QCD Lagrangean, effective field theory simplifies the problem
considerably by factorising it into a short distance part (subsumed into the
coefficient of the Lagrangean) and a long distance part which contains the
infrared-relevant physics and is dealt with by effective field theory methods.
They provide an answer of finite accuracy because higher order corrections are
systematically calculable and suppressed in powers of the expansion parameter
$Q$. Hence, the power counting allows for an error estimate of the final
result, with the natural size of all neglected terms known to be of higher
order.  The power of an effective field theory is also that relativistic
effects, chiral dynamics and external currents are included systematically, and
that extensions to include e.g.~parity violating effects are straightforward.
Gauged interactions and exchange currents are unambiguous.  Results obtained
with EFT are easily dissected for the relative importance of the various terms.
Because only $S$-matrix elements between on-shell states are observables,
ambiguities nesting in ``off-shell effects'' are absent.  On the other hand,
because only symmetry considerations enter the construction of the Lagrangean,
effective field theories are less restrictive as no assumption about the
underlying QCD dynamics is incorporated.

\absatz This article presents the first systematic effective field theory
calculation of three body properties with pions.  For the theory in which pions
are integrated out, calculations of three body observables were presented in
\cite{Stooges2}, and $\mathrm{S}$ wave quartet neutron-deuteron scattering was
computed up to NNLO below the deuteron breakup threshold. We extend this
calculation above threshold and include pions explicitly.

We choose the quartet $\mathrm{S}$ wave channel of $nd$ scattering for this
investigation because it provides a laboratory in which many complications of
the other channels are not encountered: The absence of Coulomb interactions
ensures that only properties of the strong interactions are probed. The Pauli
principle forbids three body forces~\cite{Stooges2} in the first few orders.
Because the calculation is parameter-free, it allows one to determine the range
of validity of the KSW scheme without a detailed analysis of the fitting
procedure. Although the quartet scattering length is large
($1/a^{{}^4\mathrm{S}_\frac{3}{2}}=31\;\MeV$), no extra fine tuning except
the one for the deuteron is required.

\absatz The article is organised as follows: After describing the effective
theory and its power counting in Sect.~\ref{sec:efftheory}, we apply it to $nd$
scattering in Sect.~\ref{sec:threebody}, followed by a discussion of our results
in Sect.\ \ref{sec:conclusion}.


\section{The Effective Theory}
\label{sec:efftheory}

\subsection{Re-formulating the Lagrangean}
\label{sec:lagrangean}

The first terms in the most general Lagrangean satisfying the QCD symmetries
including pions and nucleons are
\begin{eqnarray}\label{ksw}
   \mathcal{L}_{NN}&=&N^\dagger(\ii \de_0+\frac{\dev^2}{2M})N+
   \;\frac{\fpi^2}{8}\;
   \tr[(\partial_\mu \Sigma^\dagger)( \partial^\mu \Sigma)]+\nonumber\\
   &&+\;\frac{\fpi^2}{4}\;\omega \;\tr [\mathcal{M}_q (\Sigma^\dagger
   +\Sigma)]+g_A N^\dagger \vec{A}\cdot\sigma N+\\
   & &
   -\;C_0 (N^\T P^i N)^\dagger \ (N^\T P^i N)
   + \;\frac{C_2}{8}
   \left[(N^\T P^i N)^\dagger (N^\T P^i
     (\stackrel{\scriptscriptstyle\rightarrow}{\de}-
      \stackrel{\scriptscriptstyle\leftarrow}{\de})^2 N)+
   \mathrm{h.c.}\right]-\nonumber\\
   & &-\; \omega D_2 \; \tr [\mathcal{M}_\mathrm{q}] (N^\T P^i N)^\dagger\
   (N^T P^i N)
   + \dots\;\;,\nonumber
\end{eqnarray}
where $N={p\choose n}$ is the nucleon doublet of two-component spinors and
$P^i$ is the projector onto the iso-scalar-vector channel,
\begin{equation}\label{proj}
  P^{i,\,b\beta}_{a\alpha}=\frac{1}{\sqrt{8}}\; (\sigma_2\sigma^i)_\alpha^\beta
  \;(\tau_2)_a^b \;\;,
\end{equation}
$\sigma$ ($\tau$) being the Pauli matrices acting in spin (iso-spin) space.
The field $\xi$ describes the pion,
\begin{equation}\label{xi}
  \xi(x)=\sqrt{\Sigma}=e^{\ii \Pi/\fpi}\;\;,
  \qquad \Pi=\left(
  \begin{array}{cc}
    \frac{\pi^{0}}{\sqrt{2}} & \pi^{+} \\
    \pi^{-}
     &-\frac{\pi^0}{\sqrt{2}}
  \end{array}\right)\;\;.
\end{equation}
$D_\mu$ is the chiral covariant derivative $D_\mu=\partial_\mu+V_\mu$, and the
vector and axial currents are given by
\begin{equation}\label{VandA}
  V_\mu=\frac{1}{2}(\xi\partial_\mu\xi^\dagger+\xi^\dagger\partial_\mu\xi)\;\;,
  \quad
 A_\mu=\frac{\ii}{2}(\xi\partial_\mu\xi^\dagger-\xi^\dagger\partial_\mu\xi)
  \;\;. 
\end{equation}
The pion decay constant is normalised to be $\fpi=130\;\MeV$,
$\mathcal{M}_\mathrm{q}=\mathrm{diag}(m_\mathrm{u},m_\mathrm{d})$ is the quark
mass matrix, and the constant $\omega$ is chosen such that
$\mpi^2=\omega(m_\mathrm{u}+m_\mathrm{d})$. From now on, we split the
coefficient of the leading four point interaction, $C_0$, into a leading
($C_0^{(-1)}$) and a sub-leading piece ($C_0^{(0)}$) and absorb $D_2$ into
$C_0^{(0)}$ since these two terms can be distinguished only in processes
involving two or more pions sensitive to explicit chiral symmetry breaking by
the finite quark masses that will not appear here.

The Lagrangean has all the symmetries of QCD, including $SU(2)_\mathrm{L}\times
SU(2)_\mathrm{R}$ chiral invariance under which the pion and nucleon fields
transform as
\begin{equation}\label{chiral}
  \xi\rightarrow L \xi U^\dagger=U \xi R^\dagger\;\;,  \qquad
  \Sigma\rightarrow L \Sigma R^\dagger\;\;, \qquad
  N\rightarrow U N\;\;,
\end{equation}
where $L$ and $R$ are constant $SU(2)$ matrices and $U$ is a complicated,
$SU(2)$ matrix-valued function of $L,\;R$ and the pion fields.  The chiral
Lagrangean differs from the one used in Heavy Baryon Chiral Perturbation Theory
by the inclusion of four and other multi-nucleon contact forces, the latter of
which are not explicitly shown here.

\absatz We now recapitulate the scaling properties of the operators in the
Lagrangean (\ref{ksw}) necessary to establish the power counting in the next
section as suggested by Kaplan, Savage and Wise~\cite{KSW}. Because momenta
scale like $Q$ and energies like $Q^2/M$ in the non-relativistic r\'egime,
na{\ia}ve dimensional arguments suggest that the two-nucleon force term with
$n$ derivatives should scale as
\begin{equation}\label{naive}
  C_n\sim\frac{1}{M \Lambda^{n+1}}\;\;.
\end{equation}
However, the existence of unnaturally shallow bound states and large scattering
lengths indicates the presence of fine tuning.  The inverse deuteron size is
$\gamma=46\;\MeV$ and the scattering length in the nucleon-nucleon singlet
channel is $\sim 24$ fm, both of which are certainly not typical QCD scales.
Most likely, these small scales do not arise from the fact that the real world
is close to the chiral limit: In the singlet channel, for instance, the one
pion exchange potential vanishes in the chiral limit and thus cannot be the
cause of the fine tuning. The fine tuning then must be a result of short
distance physics. This, in turn, implies that a na{\ia}ve estimate of sizes of
graphs and counterterms using (\ref{naive}) may be misleading due to precise
cancellations and/or enhancements. The exact nature of these cancellations
and/or enhancements depends on the particular way the short distance physics
contribution is split between loops and counterterms and, consequently,
acquires different forms in different regularisation and subtraction schemes.
Using a cut-off scheme, for example, the sizes of the coefficients $C_n$
satisfy (\ref{naive}) with $\Lambda$ the cut-off scale, but they are fine tuned
in such a way that sums of diagrams are parametrically larger/smaller than each
diagram individually, complicating the power counting. It is also hard to keep
chiral and gauge invariance with those regulators. One convenient way of
keeping track of the new scale generated by the fine tuning was suggested by
Kaplan, Savage and Wise~\cite{KSW}. It consists in the use of dimensional
regularisation and of a subtraction scheme called Power Divergence Subtraction
(PDS), in which not only the pole present in $4$ dimensions is subtracted (as
in the minimal subtraction scheme), but also poles present in $3$ dimensions
(finite pieces in $4$ dimensions) are removed. This unusual subtraction scheme
re-shuffles a finite piece of the short distance contributions from loops to
counterterms and changes the scaling of the short distance constants from
(\ref{naive}) to
\begin{eqnarray}\label{scalingksw}
  C_0^{(-1)}&\sim&\frac{1}{M Q}\;\;,\nonumber\\
  C_0^{(0)}&\sim&\frac{1}{M \Lambda}\;\;,\\
  C_2&\sim&\frac{1}{M \Lambda Q^2}\;\;.\nonumber
\end{eqnarray}
Here, $Q$ stands for a low energy scale like the external momenta, the inverse
scattering length or the pion mass.  Other, equivalent ways of accomplishing
the same result by performing a subtraction at finite off-shell momenta were
suggested, too~\cite{MehenSteward,Gegelia}. One surprising result arises from
this analysis because chiral symmetry implies a derivative coupling of the pion
to the nucleon at leading order.  The contribution from one pion exchange
includes a factor of $Q^{-2}$ from the pion propagator and a factor of $Q^2$
coming from the pion-nucleon vertices, so that for momenta of the order of the
pion mass, the one pion exchange scales as $Q^0$ and is {\it smaller} than the
contact piece $C_0^{(-1)}$ which according to (\ref{scalingksw}) scales as
$Q^{-1}$. Iterated pion exchanges are suppressed even further. On the other
hand, repeated iterations of the $C_0^{(-1)}$ term are not suppressed, and the
two-nucleon system is described at leading order by an infinite number of
iterations of the leading contact interaction, as will be demonstrated below.
Pion exchange and higher derivative contact terms appear hence only as
perturbations at higher orders. In this scheme, the only non-perturbative
physics responsible for nuclear binding is extremely simple, and the more
complicated pion contributions are at each order given by a finite number of
diagrams. In the two body sector, this simplification allows for the derivation
of simple analytic expressions of deuteron observables. In the three body
sector discussed in this paper, even the leading order calculation will turn
out to be too complex for a fully analytical solution. Still, the equations
that need to be solved are computationally trivial and can furthermore be
improved systematically by higher order corrections that involve only (partly
analytical, partly numerical) integrations, as opposed to many-dimensional
integral equations arising in other approaches.

The estimate of the three-nucleon force terms is also influenced by the fine
tuning in the two-nucleon sector. In the quartet channel though, where the spin
of the three participating nucleons are aligned, the Pauli principle forbids a
contribution from the contact three body force without derivatives.  The
remaining ones are suppressed and do not contribute to the order we are working
here.

\absatz For our purposes, it is convenient to use a Lagrangean equivalent to
(\ref{ksw}) containing an additional field $d^i$ carrying the quantum numbers
of the deuteron, such that the four nucleon contact interactions are removed:
\begin{eqnarray}\label{dlag}
   \mathcal{L}_{Nd}&=&N^\dagger (\ii \partial_0 +\frac{\dev^2}{4 M})N+ d^{i
  \dagger} \left[w(\ii \partial_0 +\frac{\nabla^2}{4
  M})-\Delta^{(-1)}-\Delta^{(0)}\right]d^i+\nonumber\\
   & &+\;y\left[d^{i \dagger} (N^\T P^i N) + \mathrm{h.c.}\right] + \dots\;\;,
\end{eqnarray}
\noindent where
\begin{eqnarray}\label{ytoC}
  w&=&-1\;\;,\nonumber\\
  y^2&=&\frac{(C_0^{(-1)})^2}{M C_2}\;\;,\nonumber\\
  \Delta^{(-1)}&=&-\frac{C_0^{(-1)}}{M C_2}\;\;,\\
  \Delta^{(0)}&=&\frac{C_0^{(0)}}{M C_2}\;\;,\nonumber
\end{eqnarray}
\noindent and the pion nucleon interactions are still given by the
terms of (\ref{ksw}). Analogously to $C_0$, we split $\Delta$ into a leading
($\Delta^{(-1)}$) and a sub-leading piece ($\Delta^{(0)}$). The ``wrong'' sign
in the kinetic term for $w=-1$ does not spoil unitarity as can be seen from the
equivalence of (\ref{ksw}) and (\ref{dlag}) proven below. Using (\ref{ytoC})
and the scaling of the constants $C_0^{(-1)},\;C_0^{(0)}$ and $C_2$ in the KSW
scheme, we find
\begin{eqnarray}\label{deltascaling}
\Delta^{(-1)}&\sim&\frac{Q \Lambda}{M}\;\;,\nonumber\\
 \Delta^{(0)}&\sim&\frac{Q^2}{M}\;\;,\\
 y^2&\sim&\frac{\Lambda}{M^2}\;\;\nonumber.
\end{eqnarray}
The Lagrangean in (\ref{dlag}) is {\it not} the most general one involving
nucleons, pions and ``deuteron'' fields.  Since this Lagrangean contains more
fields than the minimum number necessary, there is a large reparametrisation
invariance that can be used to choose the coefficients of some terms to vanish.
This freedom in rewriting equivalent Lagrangeans will be used later to
accomplish non-trivial rearrangements of diagrams in order to simplify
computations. To see the equivalence of (\ref{dlag}) and (\ref{ksw}), we simply
perform the Gau\3ian integration over the field $d^i$ in the path integral.
After eliminating the terms containing time derivatives by a field
re-definition
\begin{eqnarray}\label{fieldredef}
  N&\rightarrow& N+\frac{y^2}{8\ (\Delta^{(-1)})^2}\; P^{i \dagger}
  N^\dagger (N^\T P^i N)\;\;,\nonumber\\
 N^\dagger&\rightarrow& N^\dagger+\frac{y^2}{8\ (\Delta^{(-1)})^2}\;
  (N^\T P^i N)^\dagger P^{i}N\;\;,
\end{eqnarray}
and using (\ref{ytoC}) we are left with (\ref{ksw}) plus i) terms of higher
order, ii) a three-nucleon force term with no derivatives that does not
contribute to the quartet scattering, and iii) multi-nucleon forces not
relevant to the three-nucleon system.  Because all those contributions can be
absorbed into terms present already in (\ref{ksw}) the two formulations of the
theory are therefore indeed equivalent. The choice of (\ref{dlag}) itself is of
course not unique. For example, the kinetic term of the deuteron field could
have been dropped in favour of keeping the $C_2$ term of the nuclear four point
interaction. We choose this form for practical reasons related to the numerical
integration, as will become apparent later.

\subsection{Power Counting and Parameter Determination}
\label{sec:powercounting}

We will briefly state the power counting used here and discuss a couple of
examples that will be relevant later. More details can be found in~\cite{KSW}.
We need to determine the scaling of different contributions as the external
momenta $p$, the inverse scattering length $ a^{{}^3S_1}$ and the pion mass go
to zero in order to estimate their relative importance. We will take $p\sim 1/
a^{{}^3S_1}\sim m_\pi$ and denote this scale by $Q$. Our expansion will be in
powers of $Q/\Lambda\sim p/\Lambda\sim 1/( a^{{}^3S_1}\Lambda)\sim
m_\pi/\Lambda$, but powers of $p\,a^{{}^3S_1}\sim m_\pi a^{{}^3S_1}\sim 1$ will
be kept to all orders.

\begin{figure}[!htb]
  \begin{center}
    \feynbox{40\unitlength}{
            \begin{fmfgraph*}(40,40)
              \fmfleft{i} \fmfright{o} \fmf{double,width=thin}{i,o}
            \end{fmfgraph*}}
          \hq$=$\hq \feynbox{40\unitlength}{
            \begin{fmfgraph*}(40,40)
              \fmfleft{i} \fmfright{o} \fmf{vanilla,width=1.5*thick}{i,o}
            \end{fmfgraph*}}
          \hq$+$\hq \feynbox{40\unitlength}{
            \begin{fmfgraph*}(40,40)
              \fmfleft{i} \fmfright{o}
              \fmf{vanilla,width=1.5*thick,tension=5}{i,v1}
              \fmf{vanilla,width=1.5*thick,tension=5}{o,v2}
              \fmf{vanilla,width=thin,left=0.65}{v1,v2}
              \fmf{vanilla,width=thin,left=0.65}{v2,v1}
            \end{fmfgraph*}}
          \hq$+$\hq \feynbox{70\unitlength}{
            \begin{fmfgraph*}(70,40)
              \fmfleft{i} \fmfright{o}
              \fmf{vanilla,width=1.5*thick,tension=5}{i,v1}
              \fmf{vanilla,width=1.5*thick,tension=5}{v2,v3}
              \fmf{vanilla,width=1.5*thick,tension=5}{v4,o}
              \fmf{vanilla,width=thin,left=0.65}{v1,v2}
              \fmf{vanilla,width=thin,left=0.65}{v2,v1}
              \fmf{vanilla,width=thin,left=0.65}{v3,v4}
              \fmf{vanilla,width=thin,left=0.65}{v4,v3}
            \end{fmfgraph*}}
          \hq$+$\hq \feynbox{100\unitlength}{
            \begin{fmfgraph*}(100,40)
              \fmfleft{i} \fmfright{o}
              \fmf{vanilla,width=1.5*thick,tension=5}{i,v1}
              \fmf{vanilla,width=1.5*thick,tension=5}{v2,v3}
              \fmf{vanilla,width=1.5*thick,tension=5}{v4,v5}
              \fmf{vanilla,width=1.5*thick,tension=5}{v6,o}
              \fmf{vanilla,width=thin,left=0.65}{v1,v2}
              \fmf{vanilla,width=thin,left=0.65}{v2,v1}
              \fmf{vanilla,width=thin,left=0.65}{v3,v4}
              \fmf{vanilla,width=thin,left=0.65}{v4,v3}
              \fmf{vanilla,width=thin,left=0.65}{v5,v6}
              \fmf{vanilla,width=thin,left=0.65}{v6,v5}
            \end{fmfgraph*}}
          \hq$+\;\dots$
  \end{center}
  \caption{\label{fig:deuteronpropagator} \sl The deuteron propagator
    at leading order from the Lagrangean (\protect\ref{dlag}). The thick solid
    line denotes the bare propagator \protect$\frac{-\ii}{\Delta^{(-1)}}$, the
    double line its dressed counterpart.}
\end{figure}
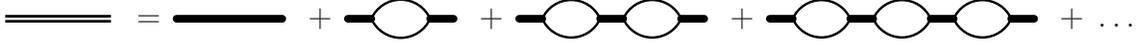

It is straightforward to estimate the loop contributions: Because the theory is
non-relativistic, the typical energies of on-shell nucleons are of the order
$Q^2/M$. Thus, the leading contribution of a diagram can be found by scaling
all momenta by a factor of $Q$ and all energies by a factor of $Q^2/M$.  The
remaining integral includes no dimensions and is taken to be of the order $Q^0$
and of natural size. This scaling implies the rule that nucleon propagators
contribute one power of $M/Q^2$ and each loop a power of $Q^5/M$. Other
r\'egimes, in which loop four momenta scale as $Q\sim\mpi$ (on-shell
propagation of pions in loops) or as $\sqrt{M\mpi}$ (pion exchanges with
momenta close to the pion production threshold), can be shown not to contribute
to the order considered here~\cite{hgpub,MehenStewardRadPions}. The vertices
provide powers of $Q$ according to (\ref{deltascaling}), implying that the
deuteron kinetic energy is sub-leading compared to the $\Delta^{(-1)}$ term.
Thus, the bare deuteron propagator is just the constant $-\ii/\Delta^{(-1)}$.
Using these rules, we find the diagrams contributing at leading order to the
deuteron propagator to be an infinite number as shown in
Fig.~\ref{fig:deuteronpropagator}, each one of the order $1/(MQ)$.
The linear divergence in each of the bubble diagrams shown in
Fig.~\ref{fig:deuteronpropagator} does not show in dimensional regularisation
as a pole in $4$ dimensions, but it does appear as a pole in $3$ dimensions
which we subtract following the PDS scheme discussed above. In other
regularisation or subtraction schemes, each term in the series depicted in
Fig.~\ref{fig:deuteronpropagator} would contribute with a different power of
$Q$, starting with the first, tree level diagram that would be of the order $
1/(M\Lambda)$. Their sum, however, will still scale as $1/(MQ)$, exemplifying
the precise cancellation discussed above, see e.g.~\cite{BiraAleph}.

The full leading order propagator $\ii\triangle^{ij}(p)$ of the deuteron field
consists hence of the geometric series shown in
Fig.~\ref{fig:deuteronpropagator},
\begin{equation}\label{dprop}
  \ii\triangle^{ij}(p) =-\;\frac{4\pi \ii}{M y^2}\;
  \frac{\delta^{ij}}{\frac{4\pi
  \Delta^{(-1)}}{M y^2} -\mu+\sqrt{\frac{\pv^2}{4}-M p_0-\ii\varepsilon}}\;\;.
\end{equation}
At leading order, all physical quantities will therefore depend on the
parameters $y$ and $\Delta^{(-1)}$ only through the combination
$C_0^{(-1)}=-y^2/\Delta^{(-1)}$, so that there is only one free parameter in
the effective theory. Either $y$ or $\Delta^{(-1)}$ can hence be chosen at
will. This is clear since the same results could have been obtained directly
from (\ref{ksw}) where only $C_0^{(-1)}$ appears. We determine this parameter
from the binding energy of the deuteron, i.e.\ from the position of the pole of
$\ii\triangle(p)$:
\begin{equation}\label{fitLO}
  C_0^{(-1)}=-\frac{y^2}{\Delta^{(-1)}}=\frac{4
  \pi}{M}\;\frac{1}{\gamma-\mu}
\end{equation}
Eq.~(\ref{dprop}) confirms the necessity to re-sum the ``bubble chain'' with
$C_0^{(-1)}$ interactions to all orders when considering momenta $p\sim \gamma$
and the scaling of $C_0^{(-1)}$ with $Q^{-1}$ when choosing as subtraction
point $\mu=\mpi\sim Q$, in agreement with the claim made previously. We will
keep $\mu=\mpi$ throughout this article. By attaching nucleon external lines to
$\ii\triangle(p)$, we obtain the nucleon-nucleon scattering amplitude which, at
this leading order, is simply the effective range expansion truncated at the
first term,
\begin{equation}\label{nn}
  \mathcal{A}^{(-1)}=-\;\frac{4\pi}{M}\;\frac{1}{\gamma+i p}\;\;.
\end{equation}
Looking at (\ref{dprop}), we derive the final rule for the power counting: Each
deuteron propagator is accompanied by a power of $1/(M y^2 Q)= M/(\Lambda Q)$.

At NLO, there are four free parameters in the Lagrangean (\ref{dlag}),
$\Delta^{(-1)},\;\Delta^{(0)},\;y$ and $w$, but only three in the equivalent
Lagrangean (\ref{ksw}).  Therefore, one of the parameters is again arbitrary.
We choose to fix the only dimensionless one, $w=-1$, giving the deuteron a
kinetic energy term with a supposedly wrong sign. Still, because the two
Lagrangeans are equivalent, unitarity and causality are not violated. The
splitting of $\Delta=\Delta^{(-1)}+\Delta^{(0)}$ between $\Delta^{(-1)}$ and
$\Delta^{(0)}$ is also arbitrary. We choose to keep $\Delta^{(-1)}$ the same as
at leading order, so that (\ref{fitLO}) is still valid at NLO and higher order
calculations will not necessitate a re-fitting of lower order
coefficients~\cite{RupakShoresh}. The remaining piece, $\Delta^{(0)}$, is then
parametrically smaller (as indicated by (\ref{deltascaling})) and is included
perturbatively. In a NNLO calculation, $C_2$ would be split in a likewise
fashion, with $C_2^{(0)}$ still given by the expression below, $C_2^{(1)}$ to
be determined anew, and so on.

The two extra conditions we use to determine the constants $\Delta^{(0)}$ and
$y$ are that the position of the deuteron pole $\gamma$, which is already
reproduced at LO by $\Delta^{(-1)}$, is unchanged, and that the nucleon-nucleon
scattering length in the triplet channel, $a^{{}^3\mathrm{S}_1}$, is
reproduced.

The condition for the pole position (Fig.~\ref{fig:polecondition}) implies with
$w=-1$ and $\mu=\mpi$ (See the later, full calculation of the deuteron
corrections at NLO in Sect.\ \ref{sec:threebody}.)
\begin{equation}\label{polecondition}
  \Delta^{(0)}=\frac{\gamma^2}{M}-\frac{y^2g_A^2\mpi^2}{32\pi^2
  \fpi^2}\left[
  (\gamma-\mpi)^2+\mpi^2\left(\ln(1+\frac{2\gamma}{\mpi})-1\right)\right]\;\;.
\end{equation}
The first term in brackets comes from the $\delta$ function part of the pion
exchange, the second one from the Yukawa part.

\begin{figure}[!htb]

  \vspace*{3ex}
  
  \begin{center}
    $ \left[ \feynbox{40\unitlength}{
            \begin{fmfgraph*}(40,20)
              \fmfleft{i} \fmfright{o} \fmf{double,width=thin}{i,v1,o}
              \fmfblob{8*thick}{v1}
            \end{fmfgraph*}}
        \right]^{-1}\bigg|_{p_0=-\frac{\gamma^2}{M},\,\pv=0}\stackrel{!}{=}0$
        \hq:\hq\hq\\[3ex]
        $ \left[ \feynbox{40\unitlength}{
            \begin{fmfgraph*}(40,20)
              \fmfleft{i} \fmfright{o} \fmf{double,width=thin}{i,o}
            \end{fmfgraph*}}
        \right]^{-1}\bigg|_{p_0=-\frac{\gamma^2}{M},\atop\,\pv=0}
        \stackrel{!}{=}0$ \hq\hq,\hq\hq $ \left[\hq \feynbox{40\unitlength}{
            \begin{fmfgraph*}(40,40)
              \fmfleft{i} \fmfright{o} \fmf{double,width=thin}{i,v1,o}
              \fmfv{decoration.shape=cross,decor.size=6*thick, label=$\fs\ii
                w(p_0-\frac{\pv^2}{4M})$,label.angle=90}{v1}
            \end{fmfgraph*}}
          \hq+\hq \feynbox{40\unitlength}{
            \begin{fmfgraph*}(40,40)
              \fmfleft{i} \fmfright{o} \fmf{double,width=thin}{i,v1,o}
              \fmfv{decoration.shape=circle,decor.filled=empty,
                decor.size=3*thick,
                label=$\fs-\ii\Delta^{(0)}$,label.angle=90}{v1}
            \end{fmfgraph*}}
          \hq+\hq \feynbox{60\unitlength}{
            \begin{fmfgraph*}(60,40)
              \fmfleft{i} \fmfright{o} \fmf{double,width=thin,tension=8}{i,v1}
              \fmf{double,width=thin,tension=8}{v2,o}
              \fmf{vanilla,width=thin,left=0.65}{v1,v2}
              \fmf{vanilla,width=thin,left=0.65}{v2,v1} \fmffreeze \fmffreeze
              \fmfipath{pa} \fmfiset{pa}{vpath(__v1,__v2)} \fmfipath{pb}
              \fmfiset{pb}{vpath(__v2,__v1)} \fmfi{dashes,foreground=red}{point
                1/2 length(pa) of pa -- point 1/2 length(pb) of pb}
              \end{fmfgraph*}}
            \hq \right]_{p_0=-\frac{\gamma^2}{M},\atop\,\pv=0}\stackrel{!}{=}0$
  \end{center}
  \caption{\label{fig:polecondition} \sl The first condition on the
    coefficients \protect$\Delta^{(-1)},\,\Delta^{(0)}$ and \protect$y$,
    (\ref{fitLO}/\ref{polecondition}). The dressed deuteron propagator has its
    pole at the physical binding energy at LO, and higher order corrections do
    not change its position. The cross denotes an energy insertion proportional
    to \protect$w$, the dot an insertion of \protect$\Delta^{(0)}$.}
\end{figure}

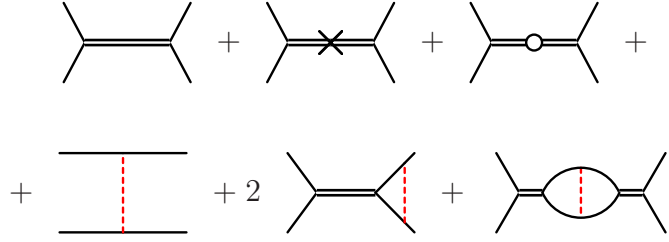
\begin{figure}[!htb]
  \begin{center}
    \feynbox{60\unitlength}{
            \begin{fmfgraph*}(60,30)
              \fmfleft{i1,i2} \fmfright{o1,o2} \fmf{vanilla,width=thin}{i1,v1}
              \fmf{vanilla,width=thin}{i2,v1} \fmf{vanilla,width=thin}{o1,v2}
              \fmf{vanilla,width=thin}{o2,v2}
              \fmf{double,width=thin,tension=0.5}{v1,v2}
            \end{fmfgraph*}}
          $+$ \feynbox{60\unitlength}{
            \begin{fmfgraph*}(60,30)
              \fmfleft{i1,i2} \fmfright{o1,o2} \fmf{vanilla,width=thin}{i1,v1}
              \fmf{vanilla,width=thin}{i2,v1} \fmf{vanilla,width=thin}{o1,v2}
              \fmf{vanilla,width=thin}{o2,v2} \fmf{double,width=thin}{v1,v3,v2}
              \fmfv{decoration.shape=cross,decor.size=6*thick}{v3}
            \end{fmfgraph*}}
          $+$ \feynbox{60\unitlength}{
            \begin{fmfgraph*}(60,30)
              \fmfleft{i1,i2} \fmfright{o1,o2} \fmf{vanilla,width=thin}{i1,v1}
              \fmf{vanilla,width=thin}{i2,v1} \fmf{vanilla,width=thin}{o1,v2}
              \fmf{vanilla,width=thin}{o2,v2} \fmf{double,width=thin}{v1,v3,v2}
              \fmfv{decoration.shape=circle,decor.filled=empty,
                decor.size=3*thick}{v3}
            \end{fmfgraph*}}
          $+$\\[5ex]
          $+$ \feynbox{60\unitlength}{
            \begin{fmfgraph*}(60,30)
              \fmfleft{i1,i2} \fmfright{o1,o2}
              \fmf{vanilla,width=thin}{i1,v1,o1}
              \fmf{vanilla,width=thin}{i2,v2,o2} \fmffreeze
              \fmf{dashes,width=thin,foreground=red}{v1,v2}
            \end{fmfgraph*}}
          $+\;2$ \feynbox{60\unitlength}{
            \begin{fmfgraph*}(60,30)
              \fmfleft{i1,i2} \fmfright{o1,o2} \fmf{vanilla,width=thin}{i1,v1}
              \fmf{vanilla,width=thin}{i2,v1}
              \fmf{vanilla,width=thin,tension=3}{o1,v3}
              \fmf{vanilla,width=thin}{v3,v2}
              \fmf{vanilla,width=thin,tension=3}{o2,v4}
              \fmf{vanilla,width=thin}{v4,v2} \fmf{double,width=thin}{v1,v2}
              \fmffreeze \fmf{dashes,width=thin,foreground=red}{v3,v4}
            \end{fmfgraph*}}
          $+$ \feynbox{80\unitlength}{
            \begin{fmfgraph*}(80,30)
              \fmfleft{i1,i2} \fmfright{o1,o2}
              \fmf{vanilla,width=thin}{i1,i0,i2}
              \fmf{vanilla,width=thin}{o1,o0,o2}
              \fmf{double,width=thin,tension=2}{i0,v1}
              \fmf{double,width=thin,tension=2}{v2,o0}
              \fmf{vanilla,width=thin,left=0.65,tension=0.3}{v1,v2}
              \fmf{vanilla,width=thin,left=0.65,tension=0.3}{v2,v1} \fmffreeze
              \fmffreeze \fmfipath{pa} \fmfiset{pa}{vpath(__v1,__v2)}
              \fmfipath{pb} \fmfiset{pb}{vpath(__v2,__v1)}
              \fmfi{dashes,foreground=red}{point 1/2 length(pa) of pa -- point
                1/2 length(pb) of pb}
              \end{fmfgraph*}}
  \end{center}
    \caption{\label{fig:NNscatteringlength} \sl \protect$NN$
      scattering at NLO gives the second condition on the coefficients
      \protect$\Delta^{(0)}$ and \protect$y$, (\ref{acondition}): The
      scattering length has the experimental value. Notation as in
      Fig.~\protect\ref{fig:polecondition}.}
\end{figure}

Imposing the experimental value of the triplet nucleon-nucleon scattering
length gives a second condition (Fig.~\ref{fig:NNscatteringlength}):
\begin{equation}\label{acondition}
  \frac{4 \pi}{M} a^{{}^3\mathrm{S}_1}=\frac{4\pi}{M}\left[\frac{1}{\gamma}+
  \frac{4\pi}{M}\frac{\Delta^{(0)}}{y^2\gamma^2}\right]
\end{equation}
Pion contributions do not appear in (\ref{acondition}) because they vanish at
zero momentum in the subtraction scheme used here, the Yukawa piece cancelling
against the ``$\delta$ function at the origin'' part.
Solving (\ref{polecondition}) and (\ref{acondition}) together with
(\ref{fitLO}) leads to
\begin{eqnarray}\label{fitNLO}
  y^2&=&\frac{4\pi}{M^2}\;\frac{\gamma}{G} \;\;,\nonumber \\
  \Delta^{(-1)}&=&\frac{(\mpi-\gamma)\gamma}{M}\;\frac{1}{G} \;\;,\\
  \Delta^{(0)}&=&\frac{\gamma^2}{M}\;\frac{\gamma a^{{}^3\mathrm{S}_1} -1}{G}
  \;\;,\nonumber
\end{eqnarray}
with
\begin{equation}\label{G}
  G= \gamma a^{{}^3\mathrm{S}_1}-1+\frac{g_A^2 M}{8 \pi \fpi^2 \gamma}
  \left[(\gamma-\mpi)^2+\mpi^2\left(\ln[1+\frac{2\gamma}{\mpi}]-1\right)
  \right]\;\;.
\end{equation}
\begin{figure}[!htb]
  \begin{center}

    \vspace*{3ex}
    
    \includegraphics*[width=0.45\linewidth]{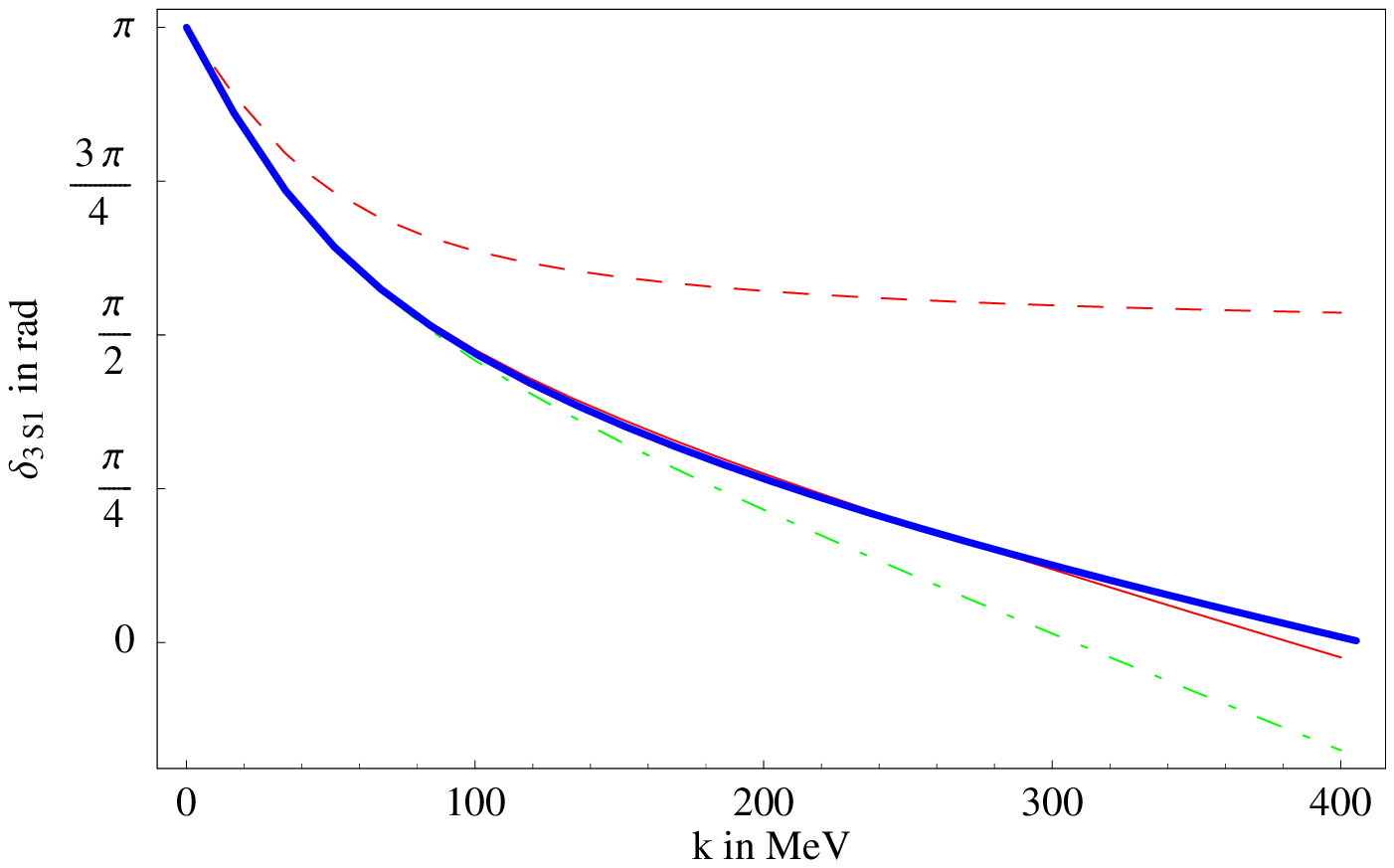}\hfill
    \includegraphics*[width=0.45\linewidth]{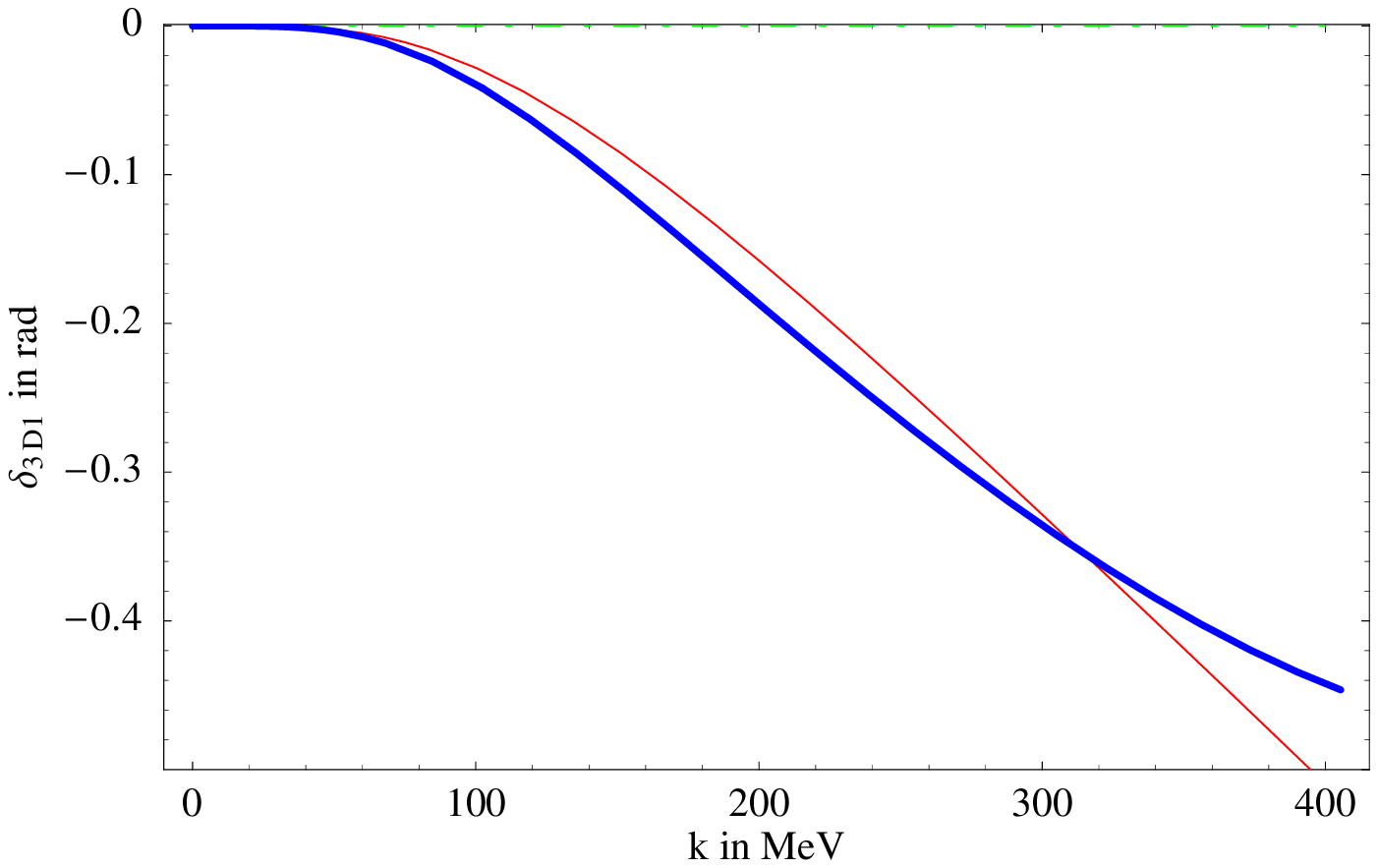}

    \vspace*{5ex}
    
    \includegraphics*[width=0.45\linewidth]{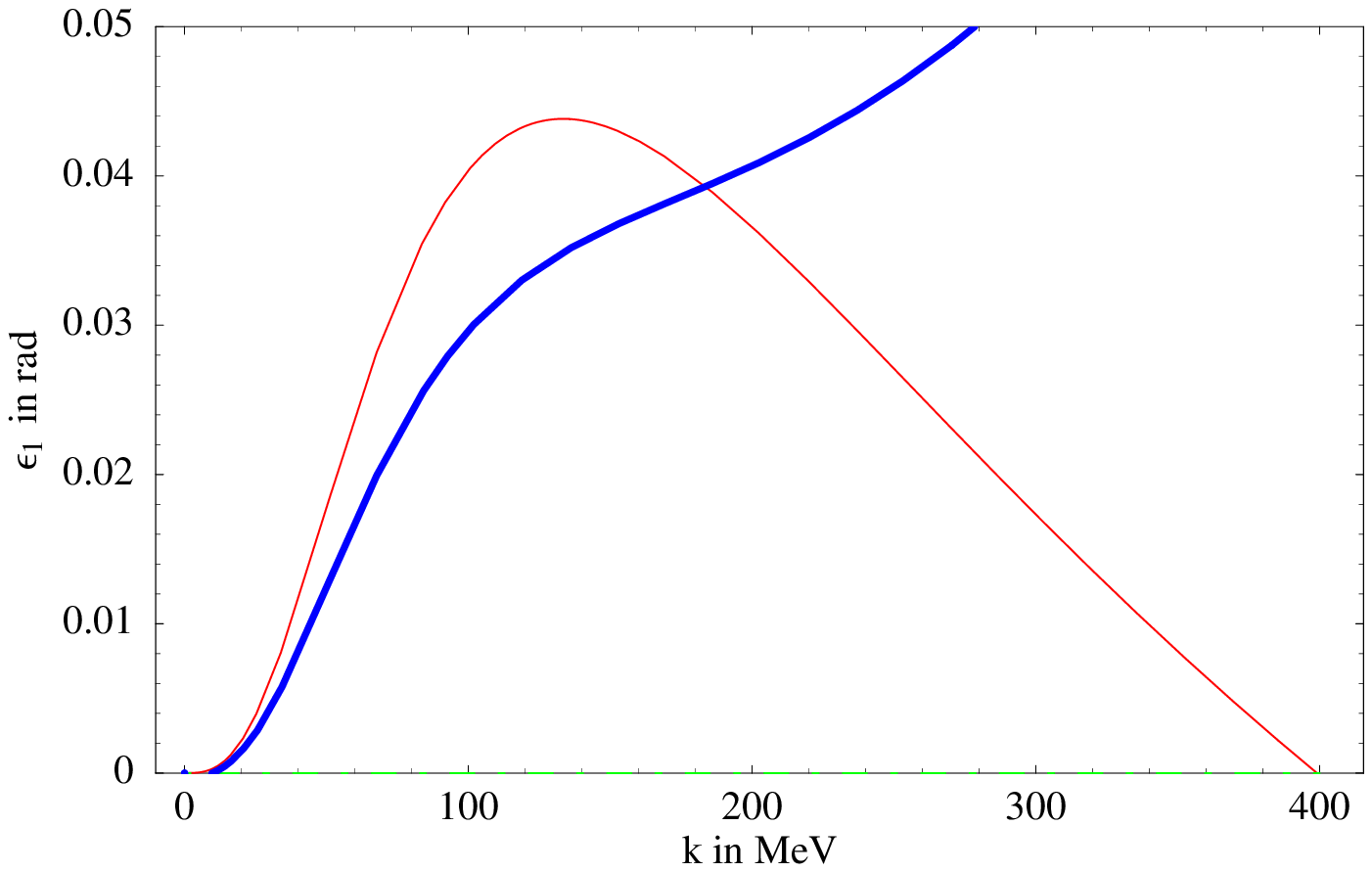}
    \caption{\label{fig:NNscatteringchannels} \sl The \protect${}^3S_1$ and
      \protect${}^3D_1$ phase shifts in the $NN$ system and the mixing angle
      \protect$\epsilon_1$ obtained with the parameters determined as described
      in the text. The dashed line is the LO result, the thin solid
      (dot-dashed) line the NLO result with perturbative pions (pions
      integrated out), and the thick solid line the Nijmegen phase shift
      analysis~\cite{Nijmegen}.}
  \end{center}
\end{figure}
Figure~\ref{fig:NNscatteringchannels} shows that the resulting phase shifts for
$NN$ scattering in the ${}^3\mathrm{S}_1-{}^3\mathrm{D}_1$ system and its
mixing parameter is in good agreement with the Nijmegen partial wave
analysis~\cite{Nijmegen}. When pions are integrated out, the phase shift of the
${}^3D_1$ channel and the ${}^3S_1-{}^3D_1$ mixing angle are identical to zero
at NLO. At LO, both quantities vanish, too.

\section{Three Body Scattering at NLO}
\label{sec:threebody}

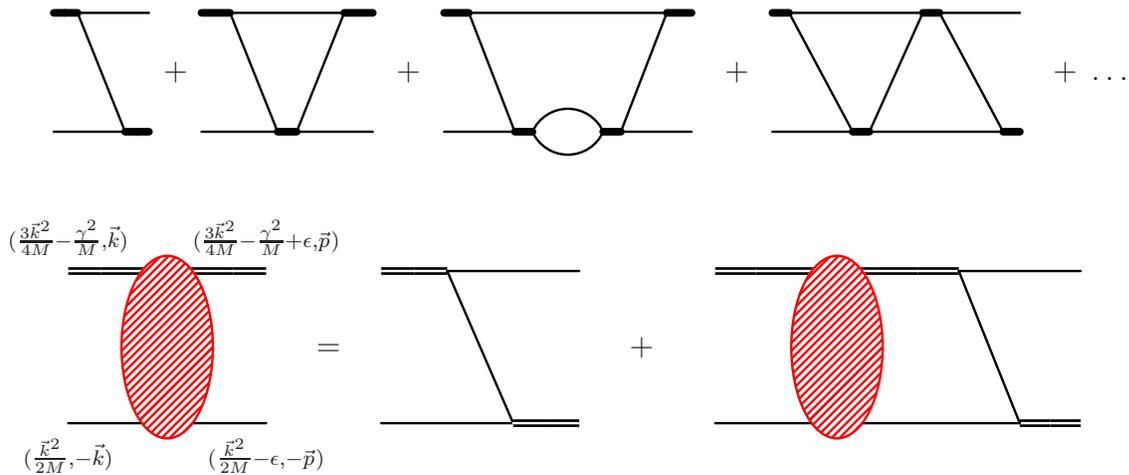
\begin{figure}[!htb]
  \begin{center}

    \vspace*{3ex}
    
    \setlength{\unitlength}{0.9pt}
    
    \feynbox{50\unitlength}{
            \begin{fmfgraph*}(50,50)
              \fmfleft{i2,i1} \fmfright{o2,o1}
              \fmf{vanilla,width=1.5*thick,tension=6}{i1,v1,v2}
              \fmf{vanilla,width=thin}{v2,o1}
              \fmf{vanilla,width=1.5*thick,tension=6}{v3,v4,o2}
              \fmf{vanilla,width=thin}{i2,v3} \fmffreeze
              \fmf{vanilla,width=thin}{v2,v3}
            \end{fmfgraph*}}
          \hqm$+$\hqm \feynbox{90\unitlength}{
            \begin{fmfgraph*}(90,50)
              \fmfleft{i2,i1} \fmfright{o2,o1}
              \fmf{vanilla,width=1.5*thick,tension=4}{i1,v1}
              \fmf{vanilla,width=thin}{v1,v2}
              \fmf{vanilla,width=1.5*thick,tension=4}{o1,v2}
              \fmf{vanilla,width=1.5*thick,tension=4}{v3,v4}
              \fmf{vanilla,width=thin}{i2,v3} \fmf{vanilla,width=thin}{v4,o2}
              \fmffreeze \fmf{vanilla,width=thin}{v2,v4}
              \fmf{vanilla,width=thin}{v3,v1}
            \end{fmfgraph*}}
          \hqm$+$\hqm \feynbox{130\unitlength}{
            \begin{fmfgraph*}(130,50)
              \fmfleft{i2,i1} \fmfright{o2,o1}
              \fmf{vanilla,width=1.5*thick,tension=8}{i1,v1}
              \fmf{vanilla,width=thin}{v1,v2}
              \fmf{vanilla,width=1.5*thick,tension=8}{o1,v2}
              \fmf{vanilla,width=1.5*thick,tension=4}{v3,v3a}
              \fmf{vanilla,width=1.5*thick,tension=4}{v4,v4a}
              \fmf{phantom}{v3a,v4a} \fmf{vanilla,width=thin}{i2,v3}
              \fmf{vanilla,width=thin}{v4,o2} \fmffreeze
              \fmf{vanilla,width=thin}{v2,v4} \fmf{vanilla,width=thin}{v3,v1}
              \fmffreeze \fmf{vanilla,width=thin,left=0.65}{v3a,v4a}
              \fmf{vanilla,width=thin,left=0.65}{v4a,v3a}
            \end{fmfgraph*}}
          \hqm$+$\hqm \feynbox{130\unitlength}{
            \begin{fmfgraph*}(130,50)
              \fmfleft{i2,i1} \fmfright{o2,o1}
              \fmf{vanilla,width=1.5*thick,tension=8}{i1,v1}
              \fmf{vanilla,width=thin}{v1,v4}
              \fmf{vanilla,width=1.5*thick,tension=8}{v4,v5}
              \fmf{vanilla,width=thin,tension=1.666}{v5,o1}
              \fmf{vanilla,width=1.5*thick,tension=8}{o2,v6}
              \fmf{vanilla,width=thin}{v6,v3}
              \fmf{vanilla,width=1.5*thick,tension=8}{v3,v2}
              \fmf{vanilla,width=thin,tension=1.666}{v2,i2} \fmffreeze
              \fmf{vanilla,width=thin}{v1,v2} \fmf{vanilla,width=thin}{v3,v4}
              \fmf{vanilla,width=thin}{v5,v6}
            \end{fmfgraph*}}
          \hqm$+\;\dots$
          \\[10ex]
          \feynbox{104\unitlength}{
            \begin{fmfgraph*}(104,64)
              \fmfleft{i2,i1} \fmfright{o2,o1}
              \fmfv{label=$\fs(\frac{3\kv^2}{4M}-\frac{\gamma^2}{M},,\kv)$,
                label.angle=90}{i1}
              \fmfv{label=$\fs(\frac{3\kv^2}{4M}-\frac{\gamma^2}{M}+\epsilon
                ,,\pv)$,label.angle=90}{o1}
              \fmfv{label=$\fs(\frac{\kv^2}{2M},,-\kv)$,label.angle=-90}{i2}
              \fmfv{label=$\fs(\frac{\kv^2}{2M}-\epsilon,,-\pv)$,
                label.angle=-90}{o2} \fmf{double,width=thin,tension=3}{i1,v1}
              \fmf{double,width=thin,tension=1.5}{v1,v3,v2}
              \fmf{double,width=thin,tension=3}{v2,o1}
              \fmf{vanilla,width=thin}{i2,v4,o2} \fmffreeze \fmffreeze
              \fmf{ellipse,foreground=red,rubout=1}{v3,v4}
              \end{fmfgraph*}}
            \hq$=$\hq \feynbox{104\unitlength}{
            \begin{fmfgraph*}(104,64)
              \fmfleft{i2,i1} \fmfright{o2,o1}
              \fmf{double,width=thin,tension=4}{i1,v1,v2}
              \fmf{vanilla,width=thin}{v2,o1}
              \fmf{double,width=thin,tension=4}{v3,v4,o2}
              \fmf{vanilla,width=thin}{i2,v3} \fmffreeze
              \fmf{vanilla,width=thin}{v2,v3}
              \end{fmfgraph*}}
            \hq$+$\hq \feynbox{192\unitlength}{
            \begin{fmfgraph*}(192,64)
              \fmfleft{i2,i1} \fmfright{o2,o1}
              \fmf{double,width=thin,tension=3}{i1,v1}
              \fmf{double,width=thin,tension=1.5}{v1,v6,v5}
              \fmf{double,width=thin,tension=3}{v5,v2}
              \fmf{vanilla,width=thin}{v2,o1} \fmf{vanilla,width=thin}{i2,v7}
              \fmf{vanilla,width=thin,tension=0.666}{v7,v4}
              \fmf{double,width=thin,tension=4}{v4,v3,o2} \fmffreeze \fmffreeze
              \fmf{vanilla,width=thin}{v4,v2}
              \fmf{ellipse,foreground=red,rubout=1}{v6,v7}
              \end{fmfgraph*}}
            
            \vspace*{4ex}
            
  \end{center}
    \caption{\label{fig:LOfaddeev} \sl The double infinite series of LO
      ``pinball'' diagrams, some of which are shown in the first line, is
      equivalent to the solution of the Faddeev equation (\ref{faddeevequation})
      shown in the second line. Notation as in
      Fig.~\protect\ref{fig:polecondition}.}
\end{figure}

Let us now look at the diagrams contributing to neutron-deuteron scattering in
the quartet channel. One might guess from the two-nucleon scattering experience
that all the graphs involving only $y$ and $\Delta^{(-1)}$ (e.g.~ the ones in
the first line in Fig.~\ref{fig:LOfaddeev}) contribute at leading order, and
the deuteron kinetic energy, pion exchanges and the $\Delta^{(0)}$ term appear
at NLO.  Indeed, following the rules discussed above, we can verify that the
tree level diagram is proportional to $y^2 M/Q^2\sim \Lambda/(MQ^2)$. The
$n$-loop diagram with a {\it bare} deuteron propagator is proportional to
$y^{2+2 n} (1/\Delta^{(-1)})^n (M/Q^2)^{1+2n}(Q^5/M)^n\sim \Lambda/(MQ^2)$ and
hence contributes at the same order. Dressing the deuteron propagator as in
Fig.~\ref{fig:deuteronpropagator} does not change the order of the diagram
since it only changes the bare propagator $-\ii/\Delta^{(-1)}\sim M/(Q\Lambda)$
by the dressed propagator $\ii\triangle(p)\sim 1/(My^2Q)\sim M/(Q\Lambda)$.
The infinite number of graphs shown in Fig.~\ref{fig:LOfaddeev} contributing at
leading order forms therefore a double infinite series.
First, all ``bubble chain'' graphs are re-summed into the leading order
deuteron propagator $\ii\triangle(p)$. Then, the kernel made of
$\ii\triangle(p)$ and the propagator of the nucleon exchanged is iterated an
arbitrary number of times. We are hence left with the task of summing all the
diagrams depicted in the first line of Fig.~\ref{fig:LOfaddeev}. In
contradistinction to the two-nucleon case, they do not form a geometric series
and cannot be summed analytically. However, one can obtain the solution
numerically from the integral equation pictorially shown in the lower line of
Fig~\ref{fig:LOfaddeev}. To derive this equation for the half off-shell
amplitude at LO, let us choose the kinematics as in Fig.~\ref{fig:LOfaddeev} in
such a way that the incoming (outgoing) deuteron line carries momentum $\kv$
($\pv$), energy $3 \kv^2/(4M) - \gamma^2/M$ ($3 \kv^2/(4M) -
\gamma^2/M+\epsilon$) and vector index $i$ ($j$).  The incoming (outgoing)
nucleon line carries momentum $-\kv$ ($-\pv$), energy $\kv^2/(2M)$ ($\kv^2/(2M)
- \epsilon$) and spinor and iso-spinor indices $\alpha$ and $a$ ($\beta$ and
$b$). Hence, the incoming deuteron and nucleon are on shell, and $\epsilon$
denotes how far the outgoing particles are off shell. We denote by $\ii
t_{\alpha a}^{ij\ \beta b}(\vec{k},\pv,\epsilon)$ the sum of those diagrams
with the kinematics above and read off from the lower line of
Fig.~\ref{fig:LOfaddeev} the integral equation for the half off-shell
amplitude:
\begin{eqnarray}
  \label{faddeevequation}
  \ii t_{\alpha a}^{ij\ \beta b}(\kv,\pv,\epsilon) &=& \frac{y^2}{2}\;
  (\sigma^j\sigma^i)_\alpha^\beta \;\delta_a^b\;
  \frac{\ii}{-\frac{\kv^2}{4}-\frac{\gamma^2}{M}+\epsilon
    -\frac{(\kv+\pv)^2}{M}+ \ii\varepsilon}+\\
  &  &+\;\frac{y^2}{2}\;
  (\sigma^j\sigma^k)_\gamma^\beta \;\delta_c^b\; \int
  \frac{d^4q}{(2\pi)^4}\   \ii t_{\alpha a}^{ik\ \gamma
    c}(\kv,\qv,\epsilon+q_0)\
  \ii\triangle(\frac{\kv^2}{4}-\frac{\gamma^2}{M}+\epsilon+q_0,\qv)\;\times
  \nonumber\\
  & &\phantom{4y^2 (\sigma^j\sigma^k)_\gamma^\beta}\times\;
  \frac{\ii}{\frac{\kv^2}{2M}-\epsilon-q_0-\frac{\qv^2}{2M}+ \ii\varepsilon}\
  \frac{\ii}{-\frac{\kv^2}{4M}-\frac{\gamma^2}{M}+2\epsilon
    +q_0-\frac{(\qv+\pv)^2}{2M}}\;\;.\nonumber
\end{eqnarray}
The integration over $q_0$ picks the pole at $q_0=(\kv^2-\qv^2)/(2M)-\epsilon+
\ii\varepsilon$. After that, we set $\epsilon=(\kv^2-\pv^2)/(2M)$, integrate
over the angle between $\kv$ and $\pv$ to project onto the $\mathrm{S}$ wave,
and set $i=(1+\ii 2)/\sqrt{2},\;j=(1-\ii 2)/\sqrt{2},\;\alpha=\beta=1,\;a=b=1$,
to pick up the spin quartet part. Denoting this projected amplitude by $\ii
t_0(k,p)$, we find
\begin{eqnarray}\label{eqfort}
  t_0(k,p)&=&-\;\frac{y^2 M}{2p k}\ln\left[\frac{p^2+p k + k^2-ME-
    \ii\varepsilon}{p^2-p k
    + k^2-ME- \ii\varepsilon}  \right]-\\
 &&-\;\frac{1}{\pi}\int\limits_0^\infty dq\; q^2\; t_0(k,q)\;
 \frac{1}{\sqrt{\frac{3 q^2}{4}-ME- \ii\varepsilon}-\gamma}\;\frac{1}{qp}\;
 \ln\left[\frac{p^2+p q + q^2-ME- \ii\varepsilon}{p^2-p q + q^2-ME-
  \ii\varepsilon} \right]\;\;,\nonumber
\end{eqnarray}
where $E=3 k^2/(4M) - \gamma^2/M$ is the total energy.  Notice that when $p=k$,
all external legs are on-shell. This equation is just the Faddeev equation for
the case of contact forces derived before by different
methods~\cite{Skornyakov}. Although the pole in the real axis due to the
deuteron propagator is regulated by the $\ii\varepsilon$ prescription and the
logarithmic singularity occurring above threshold is integrable, both cause
numerical instabilities. We used the
Hetherington--Schick~\cite{HetheringtonSchick,Amado,Book} method to numerically
solve (\ref{eqfort}). The basic idea is to perform a rotation of the variable
$q$ into the complex plane by an angle large enough in order to avoid the
singularities in and near the real axis but small enough so not to cross the
singularities of the kernel or of the solution. One can show that the
singularities of the solution are not closer to the real axis than those of the
inhomogeneous, Born term~\cite{Brayshaw}. The solution on the real axis can
then be obtained from the solution on the deformed contour by using
(\ref{eqfort}) again, now with $k$ and $p$ on the real axis and $q$ on the
contour. The computational effort then becomes trivial, and a code runs within
seconds on a personal computer.

To obtain the neutron-deuteron scattering amplitude, we have to multiply the
on-shell amplitude $t_0(k,p)$ by the wave function renormalisation constant,
\begin{eqnarray}
  \label{Z0}
T_0(k)&=&\sqrt{Z_0}\; t_0(k,k)\; \sqrt{Z_0}\;\; , \nonumber\\
 \frac{1}{Z_0}&=& \ii \frac{\partial}{\partial p_0}\;
 \frac{1}{\ii\triangle(p)} \Big|_{p_0=-\frac{\gamma^2}{M},\,\pv=0}
 \nonumber\\
 &= & \frac{M^2 y^2}{8\pi \gamma}.
\end{eqnarray}
In contradistinction to $\ii t_0(k,p)$, the scattering amplitude $\ii T_0(k)$
depends on the parameters $y$ and $\Delta^{(-1)}$ only through the observable
$\gamma$.

\absatz The power counting shows that at NLO, we have additional contributions
from: deuteron kinetic energy insertions, $\Delta^{(0)}$ insertions and pions
exchange correction to the deuteron propagator depicted in the first line of
Fig.~\ref{fig:NLOthreebody}; pionic vertex corrections to $y$ (second and third
line of Fig.~\ref{fig:NLOthreebody}); and the pion diagram of the last line of
Fig.~\ref{fig:NLOthreebody} which corrects the three particle intermediate
state (``cross diagram''). We call the first three kinds ``deuteron
insertions'' and the remaining ones ``three-body corrections'':
\begin{equation}\label{t1}
  \ii t_1(k)=\ii t_1^\mathrm{insertion}(k) + \ii t_1^\mathrm{3-body}(k)\;\;.
\end{equation}
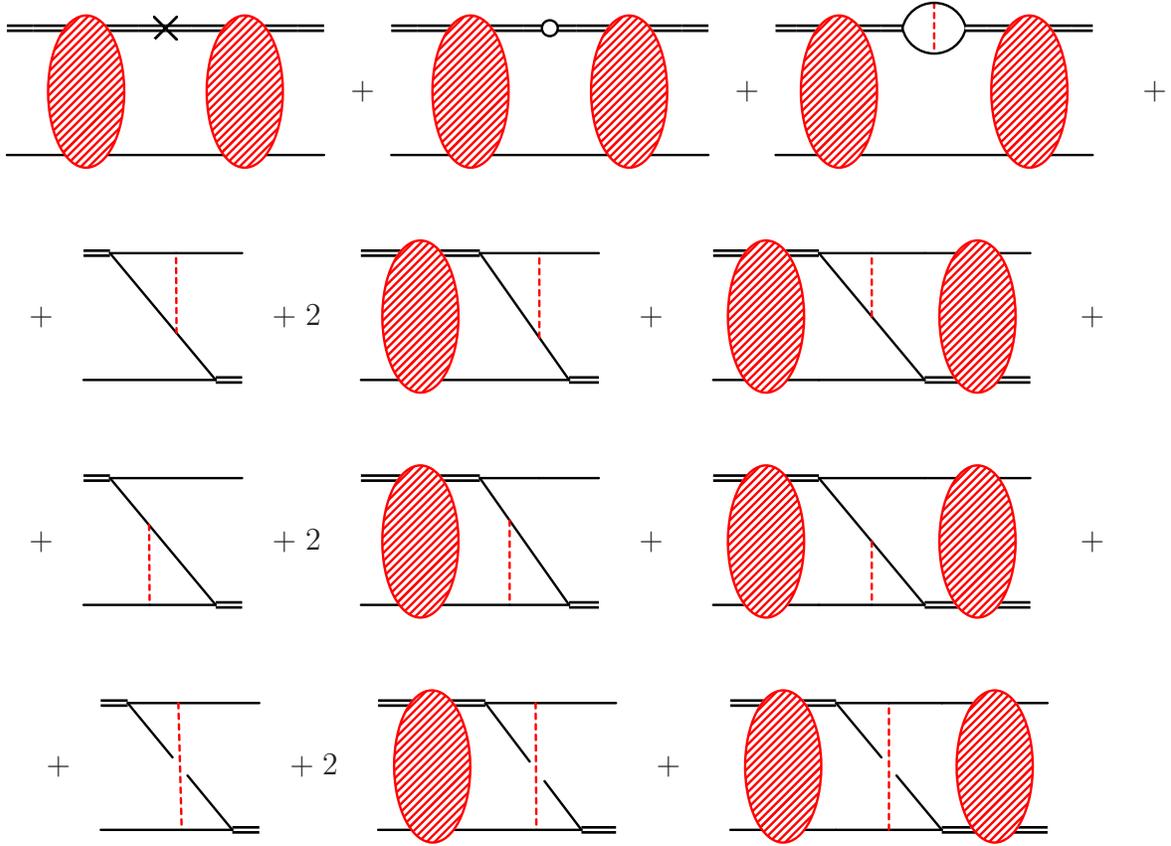
\begin{figure}[!htb]
  \begin{center}
    
    \vspace*{5ex}
    
    \setlength{\unitlength}{0.75pt}

    
    \feynbox{200\unitlength}{
            \begin{fmfgraph*}(200,64)
              \fmfleft{i2,i1} \fmfright{o2,o1}
              \fmf{double,width=thin,tension=3}{i1,v1}
              \fmf{double,width=thin,tension=1.5}{v1,v3,v2}
              \fmf{double,width=thin,tension=3}{v2,v5}
              \fmfv{decor.shape=cross,decor.size=6*thick}{v5}
              \fmf{double,width=thin,tension=3}{v5,v7}
              \fmf{double,width=thin,tension=1.5}{v7,v8,v9}
              \fmf{double,width=thin,tension=3}{v9,o1}
              \fmf{vanilla,width=thin}{i2,v4}
              \fmf{vanilla,width=thin,tension=0.5}{v4,v10}
              \fmf{vanilla,width=thin}{v10,o2} \fmffreeze \fmffreeze
              \fmf{ellipse,foreground=red,rubout=1}{v3,v4}
              \fmf{ellipse,foreground=red,rubout=1}{v8,v10}
              \fmf{vanilla,width=thin}{v3,v4} \fmf{vanilla,width=thin}{v8,v10}
              \end{fmfgraph*}}
            \hqm\hqm\hqm$+$\hqm\hqm\hqm \feynbox{200\unitlength}{
            \begin{fmfgraph*}(200,64)
              \fmfleft{i2,i1} \fmfright{o2,o1}
              \fmf{double,width=thin,tension=3}{i1,v1}
              \fmf{double,width=thin,tension=1.5}{v1,v3,v2}
              \fmf{double,width=thin,tension=3}{v2,v5}
              \fmfv{decor.shape=circle,decor.filled=empty,
                decor.size=3*thick}{v5}
              \fmf{double,width=thin,tension=3}{v5,v7}
              \fmf{double,width=thin,tension=1.5}{v7,v8,v9}
              \fmf{double,width=thin,tension=3}{v9,o1}
              \fmf{vanilla,width=thin}{i2,v4}
              \fmf{vanilla,width=thin,tension=0.5}{v4,v10}
              \fmf{vanilla,width=thin}{v10,o2} \fmffreeze \fmffreeze
              \fmf{ellipse,foreground=red,rubout=1}{v3,v4}
              \fmf{ellipse,foreground=red,rubout=1}{v8,v10}
              \fmf{vanilla,width=thin}{v3,v4} \fmf{vanilla,width=thin}{v8,v10}
              \end{fmfgraph*}}
            \hqm\hqm\hqm$+$\hqm\hqm\hqm \feynbox{200\unitlength}{
            \begin{fmfgraph*}(200,64)
              \fmfleft{i2,i1} \fmfright{o2,o1}
              \fmf{double,width=thin,tension=3}{i1,v1}
              \fmf{double,width=thin,tension=1.5}{v1,v3,v2}
              \fmf{double,width=thin,tension=3}{v2,v5}
              \fmf{vanilla,width=thin,left=0.8,tension=0.5}{v5,v6}
              \fmf{vanilla,width=thin,left=0.8,tension=0.5}{v6,v5}
              \fmf{double,width=thin,tension=3}{v6,v7}
              \fmf{double,width=thin,tension=1.5}{v7,v8,v9}
              \fmf{double,width=thin,tension=3}{v9,o1}
              \fmf{vanilla,width=thin}{i2,v4}
              \fmf{vanilla,width=thin,tension=0.333}{v4,v10}
              \fmf{vanilla,width=thin}{v10,o2} \fmffreeze \fmffreeze
              \fmf{ellipse,foreground=red,rubout=1}{v3,v4}
              \fmf{ellipse,foreground=red,rubout=1}{v8,v10}
              \fmf{vanilla,width=thin}{v3,v4} \fmf{vanilla,width=thin}{v8,v10}
              \fmffreeze \fmfipath{pa} \fmfiset{pa}{vpath(__v5,__v6)}
              \fmfipath{pb} \fmfiset{pb}{vpath(__v6,__v5)}
              \fmfi{dashes,foreground=red}{point 1/2 length(pa) of pa -- point
                1/2 length(pb) of pb}
              \end{fmfgraph*}}
            $+$
            
            \vspace*{7ex}

            
            $+$ \feynbox{100\unitlength}{
            \begin{fmfgraph*}(100,64)
              \fmfleft{i2,i1} \fmfright{o2,o1}
              \fmf{double,width=thin,tension=2.5}{i1,v1}
              \fmf{vanilla,width=thin}{v1,v3,o1}
              \fmf{double,width=thin,tension=5}{o2,v2}
              \fmf{vanilla,width=thin}{v2,i2} \fmffreeze
              \fmf{vanilla,width=thin,tension=1.7}{v2,v4}
              \fmf{vanilla,width=thin}{v4,v1} \fmffreeze
              \fmf{dashes,width=thin,foreground=red}{v4,v3}
              \end{fmfgraph*}}
            $+\;2$ \feynbox{150\unitlength}{
            \begin{fmfgraph*}(150,64)
              \fmfleft{i2,i1} \fmfright{o2,o1}
              \fmf{double,width=thin}{i1,i1a,v1}
              \fmf{vanilla,width=thin}{v1,v3,o1}
              \fmf{double,width=thin,tension=5}{o2,v2}
              \fmf{vanilla,width=thin}{v2,i2a}
              \fmf{vanilla,width=thin,tension=2.5}{i2a,i2} \fmffreeze
              \fmf{vanilla,width=thin,tension=2}{v2,v4}
              \fmf{vanilla,width=thin}{v4,v1} \fmffreeze
              \fmf{dashes,width=thin,foreground=red}{v4,v3} \fmffreeze
              \fmf{ellipse,foreground=red,rubout=1}{i1a,i2a}
              \fmf{vanilla}{i1a,i2a}
              \end{fmfgraph*}}
            $+$ \feynbox{200\unitlength}{
            \begin{fmfgraph*}(200,64)
              \fmfleft{i2,i1} \fmfright{o2,o1}
              \fmf{double,width=thin}{i1,i1a,v1}
              \fmf{vanilla,width=thin}{v1,v1a,v3}
              \fmf{vanilla,width=thin}{v3,o1a,o1}
              \fmf{double,width=thin}{v4,o2a,o2}
              \fmf{vanilla,width=thin,tension=0.5}{v4,v2}
              \fmf{vanilla,width=thin}{v2,i2a,i2} \fmffreeze
              \fmf{vanilla,width=thin}{v1,v5} \fmf{vanilla,width=thin}{v5,v4}
              \fmffreeze \fmf{dashes,width=thin,foreground=red}{v5,v1a}
              \fmffreeze \fmf{ellipse,foreground=red,rubout=1 }{i1a,i2a}
              \fmf{vanilla}{i1a,i2a} \fmf{ellipse,foreground=red,rubout=1
                }{o1a,o2a} \fmf{vanilla}{o1a,o2a}
              \end{fmfgraph*}}
            $+$
            
            \vspace{7ex}

            
            $+$ \feynbox{100\unitlength}{
            \begin{fmfgraph*}(100,64)
              \fmfleft{o1,o2} \fmfright{i1,i2}
              \fmf{double,width=thin,tension=2.5}{i1,v1}
              \fmf{vanilla,width=thin}{v1,v3,o1}
              \fmf{double,width=thin,tension=5}{o2,v2}
              \fmf{vanilla,width=thin}{v2,i2} \fmffreeze
              \fmf{vanilla,width=thin,tension=1.7}{v2,v4}
              \fmf{vanilla,width=thin}{v4,v1} \fmffreeze
              \fmf{dashes,width=thin,foreground=red}{v4,v3}
              \end{fmfgraph*}}
            $+\;2$ \feynbox{150\unitlength}{
            \begin{fmfgraph*}(150,64)
              \fmfleft{i2,i1} \fmfright{o2,o1}
              \fmf{double,width=thin}{i1,i1a,v1}
              \fmf{vanilla,width=thin}{v1,v3,o1}
              \fmf{double,width=thin,tension=5}{o2,v2}
              \fmf{vanilla,width=thin,tension=2.5}{v2,v2a}
              \fmf{vanilla,width=thin,tension=1.666}{v2a,i2a}
              \fmf{vanilla,width=thin,tension=2.5}{i2a,i2} \fmffreeze
              \fmf{vanilla,width=thin}{v2,v4}
              \fmf{vanilla,width=thin,tension=2}{v4,v1} \fmffreeze
              \fmf{dashes,width=thin,foreground=red}{v4,v2a} \fmffreeze
              \fmf{ellipse,foreground=red,rubout=1}{i1a,i2a}
              \fmf{vanilla}{i1a,i2a}
              \end{fmfgraph*}}
            $+$ \feynbox{200\unitlength}{
            \begin{fmfgraph*}(200,64)
              \fmfleft{o1,o2} \fmfright{i1,i2}
              \fmf{double,width=thin}{i1,i1a,v1}
              \fmf{vanilla,width=thin}{v1,v1a,v3}
              \fmf{vanilla,width=thin}{v3,o1a,o1}
              \fmf{double,width=thin}{v4,o2a,o2}
              \fmf{vanilla,width=thin,tension=0.5}{v4,v2}
              \fmf{vanilla,width=thin}{v2,i2a,i2} \fmffreeze
              \fmf{vanilla,width=thin}{v1,v5} \fmf{vanilla,width=thin}{v5,v4}
              \fmffreeze \fmf{dashes,width=thin,foreground=red}{v5,v1a}
              \fmffreeze \fmf{ellipse,foreground=red,rubout=1 }{i1a,i2a}
              \fmf{vanilla}{i1a,i2a} \fmf{ellipse,foreground=red,rubout=1
                }{o1a,o2a} \fmf{vanilla}{o1a,o2a}
              \end{fmfgraph*}}
            $+$
            
            \vspace*{7ex}

            
            $+$ \feynbox{100\unitlength}{
            \begin{fmfgraph*}(100,64)
              \fmfleft{i2,i1} \fmfright{o2,o1}
              \fmf{double,width=thin,tension=3}{i1,v1}
              \fmf{vanilla,width=thin,tension=1.6}{v1,v3}
              \fmf{vanilla,width=thin}{v3,o1}
              \fmf{double,width=thin,tension=3}{o2,v2}
              \fmf{vanilla,width=thin,tension=1.6}{v2,v5}
              \fmf{vanilla,width=thin}{v5,i2} \fmffreeze
              \fmf{vanilla,width=thin}{v2,v4} \fmf{phantom,tension=3}{v4,v4a}
              \fmf{vanilla,width=thin}{v4a,v1} \fmffreeze
              \fmf{dashes,width=thin,foreground=red}{v3,v5}
              \end{fmfgraph*}}         
            $+\;2$ \feynbox{150\unitlength}{
            \begin{fmfgraph*}(150,64)
              \fmfleft{i2,i1} \fmfright{o2,o1}
              \fmf{double,width=thin,tension=1.5}{i1,i1a,v1}
              \fmf{vanilla,width=thin,tension=1.6}{v1,v3}
              \fmf{vanilla,width=thin}{v3,o1}
              \fmf{double,width=thin,tension=3}{o2,v2}
              \fmf{vanilla,width=thin,tension=2.3}{v2,v5}
              \fmf{vanilla,width=thin}{v5,i2a}
              \fmf{vanilla,width=thin,tension=1.9}{i2a,i2} \fmffreeze
              \fmf{dashes,width=thin,foreground=red}{v3,v5}
              \fmf{ellipse,foreground=red,rubout=1 }{i1a,i2a}
              \fmf{vanilla}{i1a,i2a} \fmffreeze
              \fmf{vanilla,width=thin,tension=1.2}{v2,v4}
              \fmf{phantom,tension=3}{v4,v4a} \fmf{vanilla,width=thin}{v4a,v1}
              \end{fmfgraph*}}
            $+$ \feynbox{200\unitlength}{
            \begin{fmfgraph*}(200,64)
              \fmfleft{i2,i1} \fmfright{o2,o1}
              \fmf{double,width=thin}{i1,i1a,v1}
              \fmf{vanilla,width=thin}{v1,v1a,v3}
              \fmf{vanilla,width=thin}{v3,o1a,o1}
              \fmf{double,width=thin}{v4,o2a,o2}
              \fmf{vanilla,width=thin}{v4,v2a,v2}
              \fmf{vanilla,width=thin}{v2,i2a,i2} \fmffreeze
              \fmf{vanilla,width=thin}{v1,v5a}
              \fmf{phantom,width=thin,tension=3}{v5b,v5a}
              \fmf{vanilla,width=thin}{v5b,v4} \fmffreeze
              \fmf{dashes,width=thin,foreground=red}{v2a,v1a} \fmffreeze
              \fmf{ellipse,foreground=red,rubout=1 }{i1a,i2a}
              \fmf{vanilla}{i1a,i2a} \fmf{ellipse,foreground=red,rubout=1
                }{o1a,o2a} \fmf{vanilla}{o1a,o2a}
              \end{fmfgraph*}}
  \end{center}
    \caption{\label{fig:NLOthreebody} \sl The NLO contributions to $nd$
      scattering in the quartet channel. First line: Corrections to the
      deuteron propagator; second and third line: pionic vertex corrections to
      the $dNN$ vertex; fourth line: pionic corrections to three particle
      breakup in the intermediate state (``cross diagrams''). Notation as in
      Fig.~\protect\ref{fig:polecondition}.}
\end{figure}

Like the LO deuteron bubbles, the partially diverging off-shell diagrams were
calculated analytically using dimensional regularisation in order to preserve
chiral symmetry exactly at each step. The remaining integrations are finite and
hence can be treated numerically.  Each contribution in (\ref{t1}) is inserted
only once.

\absatz The insertion contributions to the NLO amplitude are given by (see the
first line of Fig.~\ref{fig:NLOthreebody}):
\begin{eqnarray}
  \label{insertion4dim}
  \ii t_1^\mathrm{insertion}(k)&=& \int \frac{d^4q}{(2\pi)^4}\;
  (\ii t_0(k,q))^2\;
  \frac{\ii}{\frac{k^2}{2M}-q_0-\frac{q^2}{2M}+\ii\varepsilon}\;\times\\
  &&\phantom{\int \frac{d^4q}{(2\pi)^4}\;}\times\;
   \left( \ii\triangle(\frac{
   k^2}{4M}-\frac{\gamma^2}{M}+q_0,\qv)\right)^2
 \;\ii\calI (\frac{k^2}{4M}-\frac{\gamma^2}{M}+q_0,\qv) \;\;,\nonumber
\end{eqnarray}
where the corrections to the deuteron are
\begin{equation}
  \label{insertion}
  \ii\calI(p_0,\pv)=\left(-\ii\Delta^{(0)}-\ii(p_0-\frac{
   \pv^2}{4M}) + \ii\Sigma_\pi(p_0,\pv)\right)
\end{equation}
with the pion contribution
\begin{eqnarray}\label{sigmapi}
  \Sigma_\pi(p_0,\pv)&=&-\frac{y^2g_A^2 M^2}{32\pi^2
    \fpi^2}\left[\calC^\mathrm{insertion}(p_0,\pv)-\mpi^2
    \calY^\mathrm{insertion}(p_0,\pv)\right]\;\;,\nonumber\\
  \calC^\mathrm{insertion}(p_0,\pv)&=&\left(\sqrt{\frac{\pv^2}{4}- M
      p_0-\ii\varepsilon}-\mpi\right)^2\;\;,\\
  \calY^\mathrm{insertion}(p_0,\pv)&=&1-\ln\left[1+
      \frac{2\sqrt{\frac{\pv^2}{4}- M
          p_0-\ii\varepsilon}}{\mpi}\right]\;\;. \nonumber
\end{eqnarray}
Here, as in the following, the pion exchange contributions separate naturally
into two pieces $\calC^\mathrm{insertion}(p_0,\pv)$ and
$\calY^\mathrm{insertion}(p_0,\pv)$ obtained by splitting the pion propagator
as
\begin{equation}\label{pion2}
  \frac{\qv^2}{\qv^2+m^2}= 1-\frac{m^2}{\qv^2+m^2}\;\;.
\end{equation}
In configuration space, the first term is a Dirac $\delta$ function resulting
in a four nucleon contact interaction, and the second is the Yukawa part. As
demonstrated at the end of this section, this simplifies the calculation
further.  The $\delta$ function contribution diverges in $3$ dimensions,
necessitating a Power Divergence Subtraction of this analytically continued pole.
The divergence in $4$ dimensions of the Yukawa piece has been removed, and the
arbitrary subtraction constant has been chosen in agreement with~\cite{KSW}
such that the total pion contribution $\Sigma_\pi(0,\vec{0})$ at zero momentum
and energy is zero. The condition (\ref{polecondition}) that the deuteron pole
is not shifted at NLO follows from (\ref{insertion}), as do the $NN$ phase
shifts of Fig.~\ref{fig:NNscatteringchannels}.

The $q_0$ integration in (\ref{insertion4dim}) picks the nucleon pole, the
angular integration is trivial, and we are left with a one dimensional integral
\begin{eqnarray}\label{insertion2}
   \ii t_1^\mathrm{insertion}(k)&=&\int\limits_0^\infty
   \frac{dq}{2\pi^2}\;q^2\;(\ii t_0(k,q))^2
   \left[ \ii\triangle(\frac{3
   k^2}{4M}-\frac{\gamma^2}{M}-\frac{q^2}{2M},\qv)\right]^2\;\times\\
   &&\phantom{\int\limits_0^\infty
   \frac{dq}{2\pi^2}}\;\times\;
   \ii\calI(\frac{3k^2}{4M}-\frac{\gamma^2}{M}-\frac{q^2}{2M},\qv)\nonumber
   \;\;,
\end{eqnarray}
which has to be performed numerically since we know $t_0(k,p)$ only
numerically.

\absatz The computation of the ``three-body'' diagrams is more involved because
a one loop pion graph has to be calculated with full off-shell kinematics and
then inserted between two half off-shell LO amplitudes, resulting after
projections in two one dimensional integrals.  We were able to perform the
$\mathrm{S}$ wave projection analytically in only part of the graphs (the ones
in the second and third line of Fig.~\ref{fig:NLOthreebody}) and we deem it
possible to perform also the other ones, but the computational advantage will
only be minimal since numerical integrations have to be performed anyway.
Thus, we are left with up to three numerical integrals, two over the magnitude
of the momenta of the two loops and one over an angle in the diagrams of the
last line of Fig.~\ref{fig:NLOthreebody}.

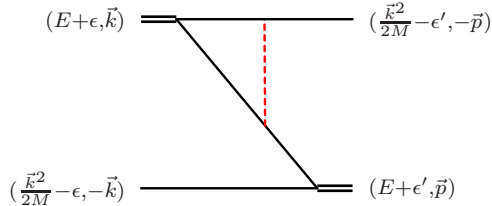
\begin{figure}[!htb]
  \begin{center}

    \vspace*{3ex}
    
    \setlength{\unitlength}{1pt} \feynbox{100\unitlength}{
            \begin{fmfgraph*}(100,64)
              \fmfleft{i2,i1} \fmfright{o2,o1}
              \fmfv{label=$\fs(E+\epsilon,,\kv)$,label.angle=180}{i1}
              \fmfv{label=$\fs(E+\epsilon^\prime,,\pv)$,label.angle=0}{o2}
              \fmfv{label=$\fs(\frac{\kv^2}{2M}-\epsilon,,-\kv)$,
                label.angle=180}{i2}
              \fmfv{label=$\fs(\frac{\kv^2}{2M}-\epsilon^\prime,,-\pv)$,
                label.angle=0}{o1} \fmf{double,width=thin,tension=2.5}{i1,v1}
              \fmf{vanilla,width=thin}{v1,v3,o1}
              \fmf{double,width=thin,tension=5}{o2,v2}
              \fmf{vanilla,width=thin}{v2,i2} \fmffreeze
              \fmf{vanilla,width=thin,tension=1.7}{v2,v4}
              \fmf{vanilla,width=thin}{v4,v1} \fmffreeze
              \fmf{dashes,width=thin,foreground=red}{v4,v3}
              \end{fmfgraph*}}
  \end{center}
  \caption{\label{fig:threebodykinematics} \sl Choice of kinematics for the
    analytical off-shell calculation of the ``three-body'' diagrams,
    exemplified at one of the pionic vertex corrections.}
\end{figure}

As depicted in Fig.~\ref{fig:threebodykinematics}, we choose the four-momentum
of the incident deuteron/nucleon as
$(E+\epsilon,\kv)$/$(\frac{\kv^2}{2M}-\epsilon,-\kv)$, and of the outgoing
deuteron/nucleon as
$(E+\epsilon^\prime,\pv)$/$(\frac{\kv^2}{2M}-\epsilon^\prime,-\pv)$. The
parameter $\epsilon$ ($\epsilon^\prime$) parametrises the off-shellness of the
initial (final) state when one chooses $E=\frac{\kv^2}{4M}-\frac{\gamma^2}{M}$.
The full off-shell behaviour of the one loop pion graphs, projected on the
quartet $\mathrm{S}$ wave state, is then given as
\begin{eqnarray}
  \calT(k,p,\epsilon,\epsilon^\prime,E)&=&
  \calT^\mathrm{vertex}(k,p,\epsilon,\epsilon^\prime,E)\;+\nonumber\\
  &&+\;\calT^\mathrm{vertex}(p,k,\frac{p^2-k^2}{2M}+\epsilon^\prime,
               \frac{p^2-k^2}{2M}+\epsilon,E-\frac{p^2-k^2}{2M})\;+\\
  &&+\;\calT^\mathrm{cross}(k,p,\epsilon,\epsilon^\prime,E)\;\;.\nonumber
\end{eqnarray}
The off-shellness of the graphs in the third line of
Fig.~\ref{fig:NLOthreebody} follows by the kinematic replacements indicated.

The first graph (second line of Fig.~\ref{fig:NLOthreebody}) is again split
into a contact and a Yukawa piece with the result
\begin{eqnarray}
  \label{pionvertexcorr}
  \calT^\mathrm{vertex}(k,p,\epsilon,\epsilon^\prime,E)&=&
  \frac{y^2 g_A M^2}{2\pi\fpi^2}\left[
  \calC^\mathrm{vertex}(k,p,\epsilon,\epsilon^\prime,E)-\mpi^2
  \calY^\mathrm{vertex}(k,p,\epsilon,\epsilon^\prime,E)\right]\;\;,\nonumber\\
  \calC^\mathrm{vertex}(k,p,\epsilon,\epsilon^\prime,E)&=&
  \frac{1}{kp}\left(\calP-\mpi\right)
  \ln\left[
    \frac{\calK^2+\calD+(p-\frac{k}{2})^2}{\calK^2+\calD+(p+\frac{k}{2})^2}
  \right]\;\;,\\
  \label{yukawavertex}
  \calY^\mathrm{vertex}(k,p,\epsilon,\epsilon^\prime,E)&=&
  \frac{1}{kp}\int\limits_{|p-\frac{k}{2}|}^{p+\frac{k}{2}} dz\;
  \frac{1}{z^2+\calK^2+\calD}\;\ln\left[
    \frac{z^2+2\ii z\calK+\mpi^2-\calK^2}{-z^2+2\ii z\mpi+\mpi^2-\calK^2}
    \right]\;\;,\nonumber
\end{eqnarray}
where we introduced the convenient abbreviations
\begin{eqnarray}
  \label{abbreviations}
   \calK&=&\sqrt{\frac{k^2}{4}-M(E+\epsilon)-\ii\varepsilon}\;\;,\\
   \calP&=&\sqrt{\frac{p^2}{4}-M(E+\epsilon^\prime)-\ii\varepsilon}\;\;,\\
   \calD&=&\frac{k^2-p^2}{2}-M\epsilon^\prime\;\;.
\end{eqnarray}
Again, dimensional regularisation with Power Divergence Subtraction was
necessary for calculating the contact piece $\calC^\mathrm{vertex}$.  The
analytic answer for the $\mathrm{S}$ wave projection represented by the
integral of the Yukawa piece (\ref{yukawavertex}) is lengthy but
straightforward and hence not reported here: One uses partial integration
followed by a partial fraction of the resulting denominator, where the
$\ii\varepsilon$ prescription takes care that no cuts or poles lie in the
integration region.  The final integrals are standard and lead to logarithms
and di-logarithms. The $\mathrm{S}$ wave projection of the tensor part of the
pionic correction to $y$ vanishes.

The contribution corresponding to the full off-shell behaviour of the last line
of Fig.~\ref{fig:NLOthreebody} is finite even in $d=3$:
\begin{eqnarray}
  \label{crosseddiagram}
  \calT^\mathrm{cross}(k,p,\epsilon,\epsilon^\prime,E)&=&
  -\frac{y^2 g_A M^2}{2\pi\fpi^2}\left[
  \calC^\mathrm{cross}(k,p,\epsilon,\epsilon^\prime,E)-\mpi^2
  \calY^\mathrm{cross}(k,p,\epsilon,\epsilon^\prime,E)\right]\;\;,\nonumber\\
  \calC^\mathrm{cross}(k,p,\epsilon,\epsilon^\prime,E)&=&
  \frac{1}{kp}\left\{4(\calK+\calP)
    \ln\left[\frac{(k+p)^2+4(\calK+\calP)^2}{(k-p)^2+4(\calK+\calP)^2}\right]
    \right.\;+
    \nonumber\\
    &&\;\;\;\;\;\;\;\;\left.+\;4 z
    \arctan\left[\frac{z}{2(\calK+\calP)}\right]
    \bigg|_{z=|k-p|}^{k+p}\right\}\;\;,\\
  \calY^\mathrm{cross}(k,p,\epsilon,\epsilon^\prime,E)&=&\int\limits_0^1dy\;
  \frac{1}{kp\calR_1}\left\{\ln\left[\frac{2 (k+p)^2+\calE}{2
        (k-p)^2+\calE}\right] \right.\;+\nonumber\\
    &&\phantom{\int\limits_0^1dy\;}\left.+\ln\left[\frac{2(k+p)^2y(1-y)+4
          \calR_1\calR_2(k+p) +\calF}{2(k-p)^2y(1-y)+4
          \calR_1\calR_2(k-p) +\calF}\right]\right\}\nonumber\;\;.
\end{eqnarray}
Here, the following abbreviations are used:
\begin{eqnarray}
  \calE&=&(2-3y)k^2-(1-3y)p^2+4(\calK^2y+\calP^2(1-y))+4\mpi\calR_1+
           4\mpi^2\left(1+y(1-y)\right)\nonumber\\
  \calF&=&4(4+y-y^2)\left[\calK^2y+\calP^2(1-y)\right]+
    y(1-y)\left[4\mpi^2+3\left(k^2(2-y)+p^2(1+y)\right)\right]\nonumber\\
  \calR_1&=&\left\{(2-y)(1+y)\left[2\calK^2y+2\calP^2(1-y)+\mpi^2y(1-y)\right]
    \right.\;+\\
    &&\;\;\left.+\;\frac{3}{2}\;y(1-y)\left[k^2(2-y)+p^2(1+y)\right]
    \right\}^\frac{1}{2}
    \nonumber\\
  \calR_2(z)&=&\sqrt{4\left[\calK^2y+\calP^2(1-y)\right]+z^2 y(1-y)}\nonumber
\end{eqnarray}
Because it contains polynomials of up to fourth order in $y$, the Yukawa part
$\calY^\mathrm{cross}$ wants one numerical integration over the Feynman
parameter $y$ -- although we deem it possible yet time-consuming to find the
analytic answer.

The ``three-body'' diagrams are classified according to the number of
(numerical) convolutions with the LO half off-shell amplitude necessary:
\begin{equation}
  \ii t_1^\mathrm{3-body}(k)=\ii t_1^\mathrm{3-body,\,0}(k)+
  2\ii t_1^\mathrm{3-body,\,1}(k)+\ii t_1^\mathrm{3-body,\,2}(k)
\end{equation}
Because the three-body diagrams in the first row of Fig.~\ref{fig:NLOthreebody}
are not convoluted with the LO half off-shell amplitude,
\begin{equation}
  \label{tree}
  \ii t_1^\mathrm{3-body,\,0}(k)=
  \ii\calT(k,k,0,0,\frac{\kv^2}{4M}-\frac{\gamma^2}{M})\;\;.
\end{equation}
For the three-body diagrams in the second and third row of
Fig.~\ref{fig:NLOthreebody} which need to be convoluted numerically once and
twice with the LO half off-shell amplitude, one picks the nucleon pole in each
of the integration over the loop energies to obtain
\begin{eqnarray}
  \label{oneloop}
  \ii t_1^\mathrm{3-body,\,1}(k)&=&\int\limits_0^\infty
  \frac{dq}{2\pi^2}\;q^2\;\ii t_0(k,q)\;
  \ii\calT(k,q,0,\frac{k^2-q^2}{2M},\frac{\kv^2}{4M}-\frac{\gamma^2}{M})
  \;\times\\ 
  &&\phantom{\int\limits_0^\infty\frac{dq}{2\pi^2}}
  \;\times\;\ii\triangle(\frac{3
   k^2}{4M}-\frac{\gamma^2}{M}-\frac{q^2}{2M},\qv)\;\;,\nonumber\\
  \ii t_1^\mathrm{3-body,\,2}(k)&=&\int\limits_0^\infty
  \frac{dq}{2\pi^2}\;q^2\;
  \frac{dp}{2\pi^2}\;p^2\;\ii t_0(k,q)\;\ii t_0(k,p)\;\times\nonumber\\
  &&\phantom{\int\limits_0^\infty\frac{dq}{2\pi^2}}
  \;\times\;\ii\calT(p,q,0,\frac{p^2-q^2}{2M},\frac{3\kv^2}{4M}-\frac{p^2}{2M}-
                    \frac{\gamma^2}{M})\;\times\\
  &&\phantom{\int\limits_0^\infty\frac{dq}{2\pi^2}}
  \;\times\;
  \ii\triangle(\frac{3k^2}{4M}-\frac{\gamma^2}{M}-\frac{q^2}{2M},\qv)\;
  \ii\triangle(\frac{3k^2}{4M}-\frac{\gamma^2}{M}-\frac{p^2}{2M},\pv)\;\;.
  \nonumber
\end{eqnarray}

\absatz The wave function renormalisation constant $Z$ at NLO is found from
\begin{eqnarray}
\frac{1}{Z}&=&\frac{1}{Z_0+Z_1}\nonumber\\
&\simeq&\frac{1}{Z_0}-\frac{Z_1}{Z_0^2}\nonumber\\
&=&\frac{1}{Z_0}+\frac{1}{Z_0^2}\;\ii\;\frac{\partial}{\partial
p_0}\; \ii\calI(p_0,\pv)\Big|_{p_0=\frac{\gamma^2}{M},\,\pv=0}
\end{eqnarray}
as
\begin{equation}\label{Z1}
  Z_1=Z_0^2\left[1+\frac{y^2 g_A^2 M^3}{32 \pi^2\fpi^2}\;
    \frac{\mpi-2\gamma}{\mpi+2\gamma}\right]\;\;.
\end{equation}
The NLO amplitude is therefore given by
\begin{eqnarray}
T(k)&=&Z\ t(k,k)\nonumber\\ &\simeq& (Z_0+Z_1)
(t_0(k,k)+t_1(k,k))\nonumber\\ &\simeq& T_0(k)+ Z_0 t_1(k,k)+Z_1
t_0(k,k)=T_0(k)+T_1(k)\;\;.
\end{eqnarray}
We extract $k\cot\delta(k)$ from
\begin{equation}\label{kcotg}
  T(k)\simeq T_0(k)+T_1(k)=\frac{3\pi}{M}\;\frac{1}{k\cot\delta(k) - ik}
\end{equation}
by expanding $k \cot\delta(k)$ and keeping only linear terms, i.e.
\begin{eqnarray}
  \label{kcotdeltaextraction}
  k\cot\delta&=&\ii k+\frac{3\pi\ii}{M}
  \left(\frac{1}{T_0(k)}-\frac{T_1(k)}{T_0(k)^2} \right)\;\;.
\end{eqnarray}
The phase shift $\delta(k)$ follows from solving (\ref{kcotdeltaextraction}).
Different ways of determining $\delta$ from the amplitude yield results which
are perturbatively close where the NLO correction to the amplitude is
parametrically small against the LO amplitude, i.e.\ below about $k\approx
200\;\MeV$.

\absatz Finally, we can avoid the computation of the contact pieces of the pion
contributions in the diagrams of Fig.~\ref{fig:NLOthreebody} by exploring the
reparametrisation invariance of the Lagrangean (\ref{dlag}) in order to get
rid of the contact piece of the pion exchange by adding and subtracting a term
in (\ref{ksw}),
\begin{equation}\label{addandsub}
  \mathcal{L}_{NN}\rightarrow  \mathcal{L}_{NN} +
  \frac{g_A^2}{16 \fpi^2} \;(N^\T P^i  N)^\dagger  (N^\T P^iN)
  - \frac{g_A^2}{16 \fpi^2}\; (N^\T P^i  N)^\dagger  (N^\T P^i N)\;\;.
\end{equation}
The part added is designed to cancel the contact part of the pion exchange and
is kept in the Lagrangean in (\ref{dlag}). The piece subtracted is reproduced
in (\ref{dlag}) by shifting the value of the constant $\Delta^{(0)}$ as
\begin{equation}\label{dlagchanged}
  \mathcal{L}_{Nd}\rightarrow
  \mathcal{L}_{Nd}(\Delta^{(0)}\rightarrow\Delta^{(0)}+
  \frac{{\Delta^{(-1)}}^2}{y^2}\;\frac{g_A^2}{2
  \fpi^2})+ \frac{g_A^2}{16 \fpi^2}\; (N^\T P^i  N)^\dagger  (N^\T P^iN)\;\;.
\end{equation}
The NLO wave function renormalisation constant $Z_1$ (\ref{Z1}) and the NLO
parameters (\ref{fitNLO}/\ref{G}) change accordingly. Again, a Gau\3ian
integration and field re-definition shows that the Lagrangean in
(\ref{dlagchanged}) is equivalent to the one in (\ref{addandsub}).

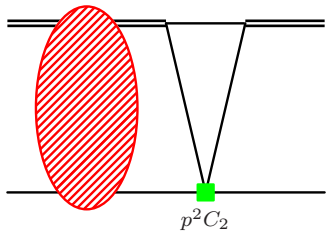
\begin{figure}[!htb]
  \begin{center}

    \vspace*{5ex}
    
    \setlength{\unitlength}{1pt} \feynbox{150\unitlength}{
            \begin{fmfgraph*}(150,64)
              \fmfleft{i2,i1} \fmfright{o2,o1}
              \fmf{double,width=thin}{i1,i1a,v1} \fmf{double,width=thin}{v2,o1}
              \fmf{vanilla,width=thin}{v1,v2}
              \fmf{vanilla,width=thin,tension=1.666}{v3,i2a}
              \fmf{vanilla,width=thin,tension=1.666}{v3,o2}
              \fmf{vanilla,width=thin,tension=2.5}{i2a,i2} \fmffreeze
              \fmf{vanilla,width=thin}{v1,v3,v2}
              \fmf{ellipse,foreground=red,rubout=1}{i1a,i2a}
              \fmf{vanilla}{i1a,i2a} \fmfv{label=$\fs
                p^2C_2$,label.angle=-90,decor.shape=square,
                decor.filled=full,decor.size=3*thick,foreground=green}{v3}
              \end{fmfgraph*}}
            
            \vspace{3ex}
            
  \end{center}
  \caption{\label{fig:desastergraph} \sl One of the graphs arising in the
    formulation of the Lagrangean with $C_2$ present, having poor convergence
    properties.}
\end{figure}

\absatz At this point, a comment on our choice of the Lagrangean (\ref{dlag})
is in order, too. As mentioned above, different Lagrangeans with the fields
$d^i$ are equivalent to the Lagrangean (\ref{ksw}). One might, for example,
choose to drop the kinetic term of the deuteron field in favour of including
into (\ref{dlag}) the $C_2$ term found in (\ref{ksw}). As with the Lagrangean
(\ref{dlag}), the original Lagrangean (\ref{ksw}) is obtained when the field
$d^i$ is integrated out, showing again the equivalence of the three
Lagrangeans. In $nd$ scattering, the diagrams to be calculated would be
different even though their sum produces the same on-shell amplitude. In
particular, one graph to be calculated is the triangle with $C_2$ on the
vertex, dressed on one side by the LO solution, Fig.~\ref{fig:desastergraph}.
According to the analysis of the asymptotic behaviour of the solution of the LO
integral equation~\cite{Danilov}, $\ii t_0(k,p)\sim 1/p^{3.17\dots}$ for $p\gg
k,\;\gamma$. Consequently, the diagram of Fig.~\ref{fig:desastergraph} has an
UV behaviour of $\frac{(Q^5)^2 Q^2}{Q (Q^2)^4}\;\frac{1}{Q^{3.17}}=Q^{-0.17}$
and hence is barely convergent, making a numerical integration difficult.  In
addition, the number of diagrams to be calculated is increased considerably.
In the approach chosen here, the insertion diagrams have an UV behaviour of
$Q^{-3.33}$. The three-body diagrams which are not convoluted with the LO half
of-shell amplitude converge like $Q^{-1}$ before, and like $Q^{-3}$ after
absorption of the contact piece of the pion graphs into the definition of
$\Delta^{(0)}$. The numerically more involved diagrams of
Fig.~\ref{fig:NLOthreebody} which have to be convoluted once (twice) with the
LO half of-shell amplitude converge like $Q^{-2.17}$ ($Q^{-3.33}$) before, and
like $Q^{-4.17}$ ($Q^{-5.33}$) after the contact piece of the pion interaction
is removed.

\section{Discussion and Conclusions}
\label{sec:conclusion}

\begin{figure}[!htb]
  \begin{center}
    \includegraphics*[width=0.9\linewidth]{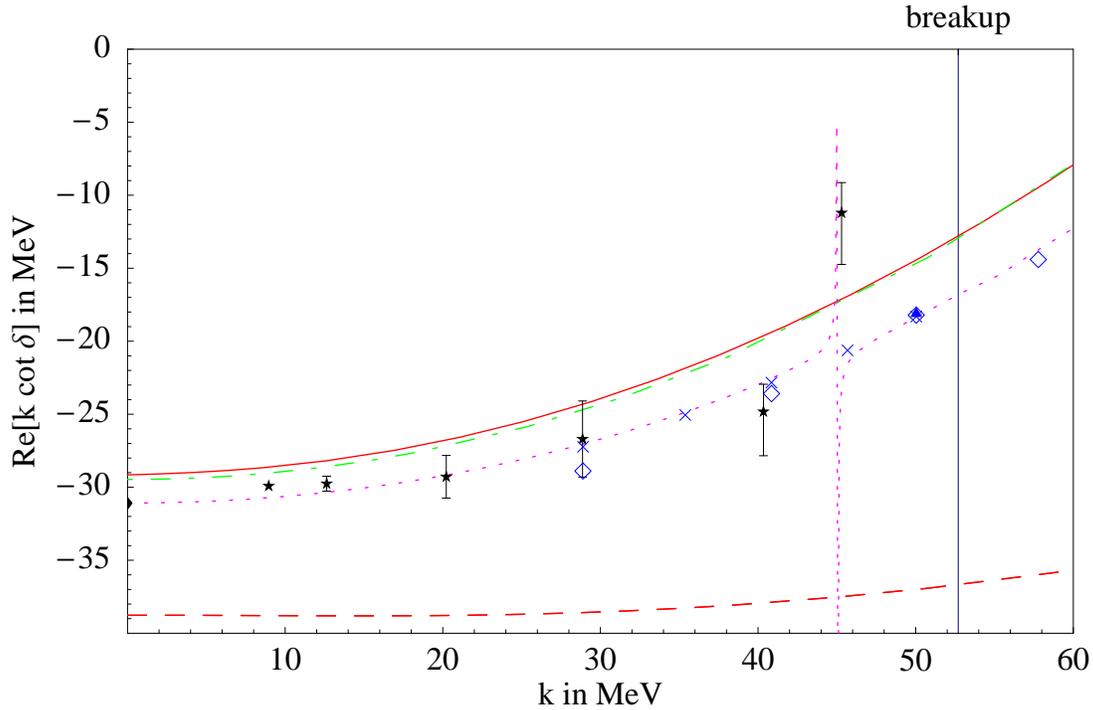}
    \caption{\label{fig:kcotdelta} \sl \protect$k\cot\delta(k)$ vs.~$k$ in the
      centre-of-mass frame below breakup and extraction of the quartet
      scattering length. The dashed line is the LO result, the solid
      (dot-dashed) line the NLO result with perturbative pions (pions
      integrated out). The dotted line is an NNLO calculation in effective
      field theory without pions~\cite{Stooges2}. The diamond at zero momentum
      is from the experiment of~\cite{Dilgetal}. Stars denote the experiment by
      van Oers and Seagrave as reported in~\cite{PhillipsBarton}. Results from
      realistic potential models are reported as squares
      from~\cite{WitalaTUNL}, crosses from~\cite{Kievsky} and triangles
      from~\cite{Hueberetal}.}
  \end{center}
\end{figure}

\begin{figure}[!htb]
  \begin{center}
    \includegraphics*[width=0.9\linewidth]{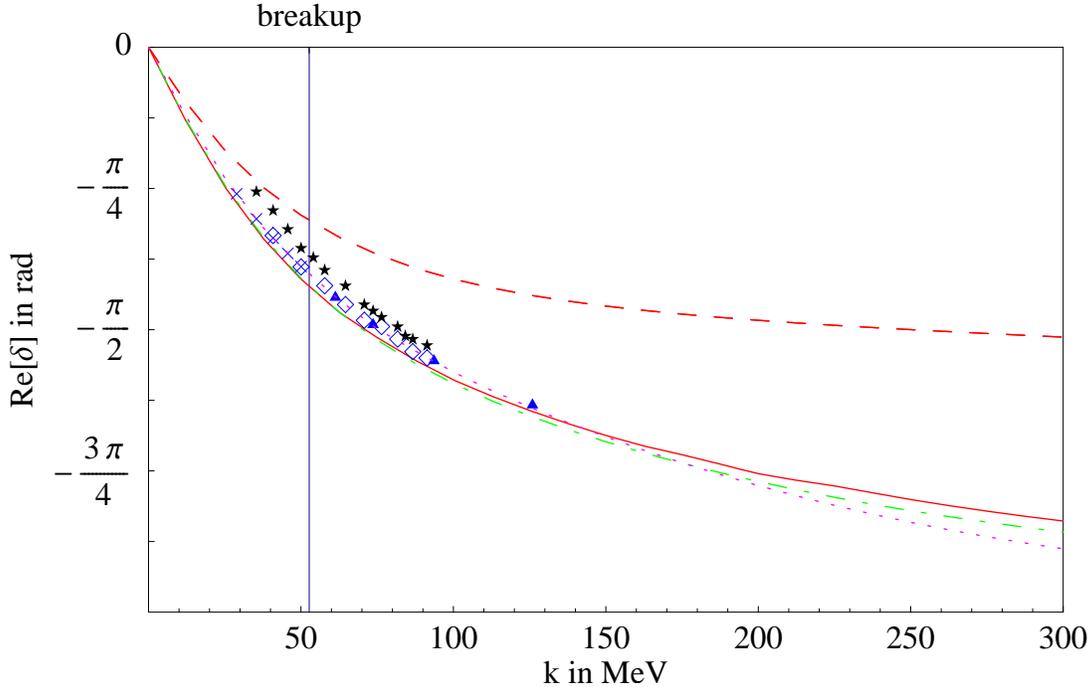}
    \caption{\label{fig:redelta} \sl The real part of the phase shift above
      breakup. The stars are taken from the TUNL $pd$ phase shift
      analysis~\cite{WitalaTUNL}. Remaining notation as in
      Fig.~\ref{fig:kcotdelta}.}
  \end{center}
\end{figure}

\begin{figure}[!htb]
  \begin{center}
    \includegraphics*[width=0.9\linewidth]{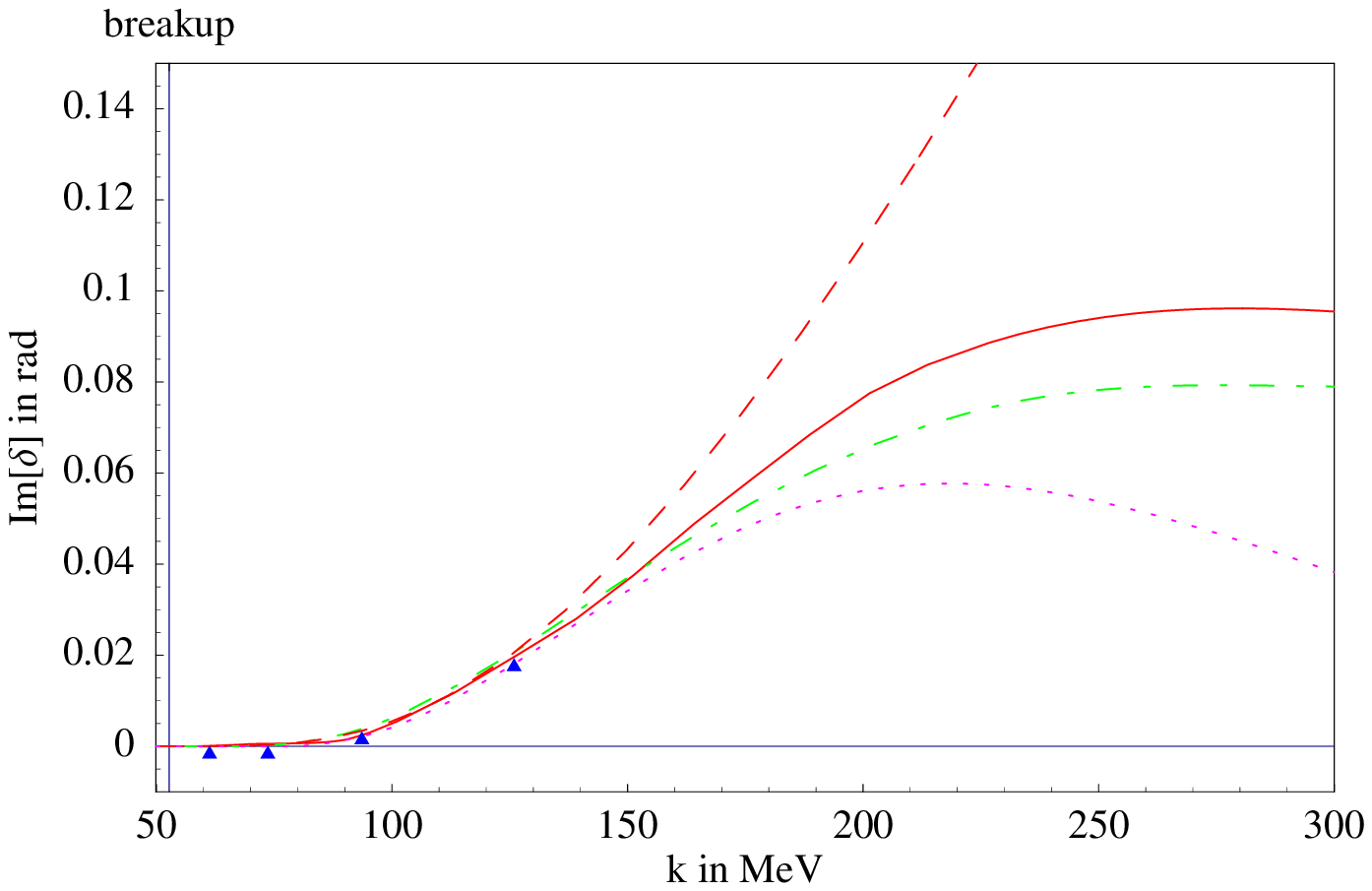}
    \caption{\label{fig:imdelta} \sl The imaginary part of the phase shift
      above breakup. Notation as in Fig.~\ref{fig:kcotdelta}.}
  \end{center}
\end{figure}

With $\hbar c=197.327\;\MeV\,\fm$, a nucleon mass of $M=938.918\;\MeV$, a
deuteron binding energy (momentum) of $B=2.225\;\MeV$ ($\gamma=45.7066\;\MeV$)
and the $NN$ triplet $\mathrm{S}$ wave scattering length
$a^{{}^3\mathrm{S}_1}=5.42\;\fm$, as well as with the physical values for the
pion parameters, $\mpi=138.039\;\MeV$, $\fpi=130\;\MeV$ and $g_A=1.25$, the
computation yields the results shown in Figs.~\ref{fig:kcotdelta},
\ref{fig:redelta} and \ref{fig:imdelta} for $k\cot\delta$ as function of the
momentum in the centre-of-mass frame below breakup and for the real and
imaginary parts of the phase shifts. Because experimental data is scarce in
this channel~\cite{PhillipsBarton} and only available below the breakup point
$k_\mathrm{breakup}=\sqrt{4/3}\;\gamma\approx 52.7\;\MeV$, we also compare to
the TUNL $pd$ phase shift analysis~\cite{WitalaTUNL} above breakup as Coulomb
and chiral symmetry breaking effects can be neglected at high momenta.  In
addition, two calculations based on the AV 18
potential~\cite{WitalaTUNL,Kievsky} and the Bonn B potential~\cite{Hueberetal}
are presented, the latter one providing the only comparison for the imaginary
part of the phase shift above breakup. Extending the results of
Ref.~\cite{Stooges2} above breakup, we also give an effective field theory
calculation to NNLO with pions integrated out. Convergence is good.


In Fig.~\ref{fig:kcotdelta}, we focus on the behaviour below threshold where
$p\ll \mpi$. In this r\'egime, the pions are expected to be integrated out
resulting in a much simpler effective (``pion-less'') theory involving contact
interactions between nucleons only~\cite{CRS,Stooges}. Besides our NLO result
with explicit pions discussed above (full line), we also show the LO, NLO and
NNLO calculation in a theory with pions integrated out (dashed, dash-dotted and
dotted line).  Simplistically, this is achieved by setting $g_\mathrm{A}=0$ in
the determination of the parameters (\ref{fitNLO}/\ref{G}) and in the
calculations of the NLO corrections. These calculations demonstrate convergence
and predict a quartet scattering length in excellent agreement with experiment.
The NLO calculation in the effective theory with and without pions give very
similar results. The parameters in the pion-less and pion-full LO and NLO
calculations were determined as described above to produce the physical values
of $\gamma$ and $a^{{}^3S_1}$, while the NNLO result taken
from~\cite{Dilgetal,Stooges} used as input $a^{{}^3S_1}$ and the effective
range $r_0=1.75\;\fm$, the difference of the two procedures to determine the
parameters being of higher order in the power counting. No free parameters
arise in any of the calculations.

The difference between the pion-full and pion-less theory should appear for
momenta of the order of $\mpi$ and higher because of non-analytical
contributions of the pion cut. However, it is very moderate for momenta of up
to $300\;\MeV$ in the centre-of-mass frame ($E_\mathrm{cm}\approx70\;\MeV$),
see Fig.~\ref{fig:redelta}.  This and the lack of experimental data makes it
difficult to assess whether the KSW power counting scheme to include pions as
perturbative increases the range of validity over the pion-less theory .
Unfortunately, numerical calculations with realistic potentials are not
available for energies considerably higher than the deuteron breakup at the
present time~\footnote{At the order we are working, an effective theory of a
  realistic potential model will be identical to the one discussed here. Hence,
  a model can serve for the present purpose.}. Still, where data or numerical
calculations are available, the pion-full theory does better than the pion-less
one.

In deuteron calculations, the numerical value of the systematic expansion
parameter $Q$ seems to be of the order $\frac{1}{3}$ at zero momentum, and we
confirm this by comparing the quartet scattering length at LO and NLO: We find
$a^{{}^4S_\frac{3}{2}}(\mathrm{LO})=(5.1\pm 1.5)\;\fm$, and at NLO with
(without) perturbative pions $a^{{}^4S_\frac{3}{2}}(\mathrm{NLO,\pi})=(6.8\pm
0.7)\;\fm$ ($a^{{}^4S_\frac{3}{2}}(\mathrm{NLO,\mathrm{no}\,\pi})=(6.7\pm
0.7)\;\fm$). At NNLO, ~\cite{Stooges} report
$a^{{}^4S_\frac{3}{2}}(\mathrm{NNLO,\mathrm{no}\,\pi})=(6.33\pm 0.1)\;\fm$, and
the experimental value is given in \cite{Dilgetal} as
$a^{{}^4S_\frac{3}{2}}(\mathrm{exp})=(6.35\pm 0.02)\;\fm$.  Comparing the NLO
correction to the LO scattering length provides one with the error estimate
quoted: The NLO calculation is estimated to be accurate up to about
$(\frac{1}{3})^2\approx 10\%$. The NLO calculations with and without pions lie
within each other's error bar. The NNLO calculation quoted above is inside the
error ascertained to the NLO calculation and carries itself an error of about
$(\frac{1}{3})^3\approx 4\%$. NLO and LO contributions become comparable for
momenta of more than $200\;\MeV$. In the imaginary part shown in
Fig.~\ref{fig:imdelta}, the same pattern emerges with a slightly more
pronounced difference between the pion-less and pions-full theory.

Because results obtained with EFT are easily dissected for the relative
importance of the various terms, we show that pionic corrections to
$nd$ scattering in the quartet $\mathrm{S}$ wave channel -- although
formally NLO -- are indeed much weaker: The calculation with perturbative pions
and with pions integrated out do not differ significantly over a wide range of
momenta. 

\absatz Future work will extend the analysis presented here to higher partial
waves and to the inclusion of Coulomb effects in order to allow for comparison
with $pd$ data.


\section*{Acknowledgements}
We are indebted to Ch.~Hanhart and D.~R.~Phillips for bringing the
Hetherington--Schick method to our attention, and to G.~Rupak, M.~J.~Savage and
the rest of the effective field theory group at the INT and the University of
Washington in Seattle for a number of valuable discussions.  It is also our
pleasure to thank the ECT*, Trento, for its hospitality during the workshop
``The Nuclear Interaction: Modern Developments'', during which this article was
finished. The work was supported in part by a Department of Energy grant
DE-FG03-97ER41014.

\newpage



\end{fmffile}
\end{document}